\documentclass[fleqn,usenatbib]{mnras}

\usepackage{newtxtext,newtxmath}

\usepackage[T1]{fontenc}

\DeclareRobustCommand{\VAN}[3]{#2}
\let\VANthebibliography\thebibliography
\def\thebibliography{\DeclareRobustCommand{\VAN}[3]{##3}\VANthebibliography}

\usepackage{graphicx}	
\usepackage{amsmath}	

\usepackage{enumitem}
\usepackage{booktabs} 

\newcommand{\maps}{MAPS}
\newcommand{\psj}{Planet. Sci. J.}

\title[Mining potential of undifferentiated asteroids]{Assessing the metal and rare earth element mining potential of undifferentiated asteroids through the study of carbonaceous chondrites}

\author[J. M. Trigo-Rodríguez et al.]{
J. M. Trigo-Rodríguez,$^{1,2}$\thanks{E-mail: trigo@ice.csic.es}
P. Grèbol-Tomàs,$^{1, 2}$
J. Ibáñez-Insa$^{3}$
J. Alonso-Azcárate$^{4}$
and M. Gritsevich$^{5, 6}$
\\
$^{1}$Institut de Ciències de l’Espai (ICE, CSIC), Campus UAB, Carrer de Can Magrans S/N, Cerdanyola del Vallès, 08193, Catalonia, Spain\\
$^{2}$Institut d’Estudis Espacials de Catalunya (IEEC), Esteve Terradas 1, Edifici RDIT, Of. 212, Parc Mediterrani de la Tecnologia (PMT),\\Campus del Baix Llobregat – UPC, Castelldefels (Barcelona), 08860, Catalonia, Spain\\
$^{3}$Geociències Barcelona (GEO3BCN - CSIC), Carrer de Lluís Sole i Sabaris, 1, Barcelona, 08028, Catalonia, Spain\\
$^{4}$Universidad de Castilla-La Mancha (UCLM), Facultad CC Ambientales y Bioquímica, Campus Fábrica de Armas, Toledo, 45071, Castilla-La Mancha, Spain\\
$^{5}$Faculty of Science, University of Helsinki, Gustaf Hallströmin katu 2, FI-00014 Helsinki, Finland\\
$^{6}$Institute of Physics and Technology, Ural Federal University, Mira str. 19, 620002 Ekaterinburg
}

\date{Accepted for publication in MNRAS on 2025 October 24. Received 2025 October 24; in original form 2025 October 03}

\pubyear{2025}

\begin{document}
\label{firstpage}
\pagerange{\pageref{firstpage}--\pageref{lastpage}}
\maketitle

\begin{abstract}
Undifferentiated asteroids, particularly the parent bodies of carbon-rich chondrite groups, might be promising candidates for future space resource utilization due to their primitive composition and potential to host valuable metals and rare earth elements. However, our understanding of their bulk elemental composition remains limited, as most data are derived from reflectance spectra with low mineralogical resolution. Sample return missions have started to change that, as returned materials are already available to study. Still the available meteorites provide a valuable source of information about the diversity of undifferentiated asteroids in the interplanetary space. To improve compositional insights, we conducted Inductively Coupled Plasma Mass Spectrometry (ICP-MS) and ICP-AES (Inductively coupled Plasma Atomic Emission Spectroscopy) analyses on a representative suite of carbonaceous chondrites. These meteorites, considered analogs of undifferentiated asteroids, preserve materials from the early solar system and provide a geochemical record of their parent bodies. Our results highlight the abundance and distribution of transition metals, siderophile elements, and rare earth elements across several chondrite groups. These findings support the view that C-type asteroids may serve as viable sources of critical materials, while also informing future mission planning, extraction strategies, and the development of new technologies for low-gravity resource operations. 
\end{abstract}




\section{Introduction}

The demand for raw materials in modern technology, including electronics, renewable energy systems, and advanced transportation, has led to significant interest in alternative sources of critical elements \citep{2019..book....DS}. Many of these elements are scarce in Earth's crust, difficult to extract, or concentrated in limited geopolitical regions. As a result, space exploration is increasingly viewed as a potential pathway for accessing new supplies of these strategic resources. Table 1 lists a selection of these critical elements along with their main technological applications. These elements are of strategic interest not only for terrestrial use but also for their potential in future space-based industries.

\begin{table*}
\centering
\small
\caption{Technologically relevant elements, compounds, applications, and relevance to undifferentiated asteroids}
\label{tab: elements}
\resizebox{\textwidth}{!}{
\begin{tabular}{@{}clp{4.2cm}p{4.8cm}p{4.8cm}@{}}
\toprule
\textbf{Symbol} & \textbf{Name} & \textbf{Relevant Compounds / Forms} & \textbf{Main Applications} & \textbf{Presence in Carbonaceous Chondrites} \\
\midrule
Li & Lithium & LiFePO\textsubscript{4}, LiCoO\textsubscript{2} & Batteries, reactor coolant, electronics & Trace amounts in CM and CI chondrites \\
Mg & Magnesium & MgO, MgSiO\textsubscript{3}, Mg\textsubscript{2}SiO\textsubscript{4}  & energy storage/battery, alloys and biomedical products & Main constituent of silicates \\
Ga & Gallium & GaAs, GaN & Semiconductors, LEDs, laser diodes & Present in CO and CM chondrites \\
Ag & Silver & Ag, AgNO\textsubscript{3} & Conductors, antibacterial coatings & Trace amounts; uncommon in CCs \\
In & Indium & ITO, InP, InGaN & LCDs, solar panels, diodes & Rarely measured in CCs \\
Au & Gold & Au, AuCl\textsubscript{3} & Electronics, spacecraft wiring, optics & Trace levels in CR and CV groups \\
Cu & Copper & Cu, CuO, Cu\textsubscript{2}O & Wiring, motors, spacecraft systems & Moderate in CO and CR chondrites \\
Zn & Zinc & ZnO, ZnS & Coatings, batteries, alloys & Abundant in CI and CM chondrites \\
C & Carbon & Graphite, organics & Composites, shielding, ISRU & Abundant in all CC groups \\
Nd & Neodymium & NdFeB magnets & Motors, turbines, spacecraft actuators & Trace REEs in CO, CV chondrites \\
Dy & Dysprosium & NdFeB + Dy alloy & High-temp magnets, EVs, defense tech & Measurable in refractory inclusions \\
La & Lanthanum & La\textsubscript{2}O\textsubscript{3} & Optics, catalysts, hydrogen storage & Trace in CAIs and fine-grained matrix \\
\bottomrule
\end{tabular}
}
\end{table*}

In scientific literature, the prevailing view is that asteroid mining should focus on metal-rich differentiated asteroids. These bodies, which likely represent fragments of larger, melted planetary precursors, are dominated by Fe-Ni metal phases and often categorized as M-type asteroids. Certainly these M-class asteroids contain very valuable contents of Fe and Ni, together with other metal-loving elements that move together with Fe-Ni alloys to the differentiated body core during the chemical segregation of planetary bodies \citep{2023P&SS..22505608C}. Metal-rich asteroids are considered, in fact, parts of bodies already melted that were destroyed in giant collisions \citep{2017E&PSL.472..152B}.

However, this work redirects attention to undifferentiated, primitive asteroids — those that have retained their primordial composition and escaped internal melting and chemical stratification. These bodies offer a unique repository of both metals and volatile-rich materials. Many of these asteroids only grew to a few hundred km in diameter and were highly porous \citep{2006ApJ...652.1768B}, so they lost a significant part of the primordial radioactive and collisional heat, preventing differentiation. As a consequence, they have preserved to some extent the primordial minerals, particularly the metal grains together with other accretionary components available when they grew from the protoplanetary disk about 4.5 Gyr ago \citep{1998cm}. For example, it has been estimated that about 16 per cent in weight of the materials that form the H group of ordinary chondrites is metallic iron \citep{2021P&SS..20605309L}. Ordinary chondrites may also be the source of other valuable metals, as recently inferred from a study focusing in Co and Cu \citep{2020P&SS..19405092L}. Our goal is to demonstrate that undifferentiated asteroids, typically overlooked in commercial mining strategies, may in fact hold substantial scientific and economic value for future exploration efforts. Importantly, they also offer access to highly fragile materials that would rarely, if ever, survive atmospheric entry to be recoverable as meteorites \citep{2009P&SS...57..243T, 2024M&PS...59.1658G}. In situ exploration or sample return is therefore the only practical means of recovering and studying these delicate constituents, which could reveal otherwise inaccessible insights into early Solar System processes \citep{2009ExA....23..785B}.

Thus, we often consider undifferentiated asteroids to be authentic survivors, but it is important to remark that some of them retained volatiles, particularly water, and were subjected to multiple collisions since their formation. As a consequence, their surfaces have been collisionally processed over the eons \citep{2016ApJ...824...12B}, so a priori we could think that the regolith covering their surfaces could be easily separated into silicates and metallic grains for In Situ Resource Utilisation (ISRU). However, in practice, lithification and breccification makes things much more complicated. In addition, different degrees of thermal annealing or even mild metamorphism is detected in carbonaceous chondrites \citep{2004GeCoA..68..673R}, leading to partial depletion of moderately volatile elements, formation of high-pressure shocked minerals, and even implantation of foreign clasts \citep{2005GeCoA..69.3419R, 2019NatAs...3..659N}. In principle, profitable extraction of Fe and Ni seems to be possible, given that metal alloys are widely present in undifferentiated asteroids, as does also occur in lunar breccias \citep{1998LPI....29.1008P}. In addition, there is an obvious interest in obtaining rare metals from these asteroids that are scarce on Earth, given the wide applications of these key materials (Table 1). It is obvious that undifferentiated asteroids that have not experienced significant aqueous alteration should contain most metals in native form, almost unaltered since accretion \citep{2014MNRAS.437..227T}. Other groups of chondrites were affected by aqueous alteration at early times after their accretion, so they often contain metals in a less profitable oxidized form \citep{1996Natur.379..701E, 2019hmep.book....4T}.

Obviously, not only metals should be considered as key asteroidal components susceptible of being of interest for mining purposes. Perhaps one of the best examples is a group of elements of difficult extraction called the Rare Earth Elements (hereafter REEs). The market for REEs is set for massive expansion, with recent forecasts pointing to an exponentially rising global demand. The REEs comprise 15 elements that have long been in electronic products and a range of other technological applications. The main REEs group comprises the 15 lanthanides: cerium (Ce), dysprosium (Dy), erbium (Er), europium (Eu), gadolinium (Gd), holmium (Ho), lanthanum (La), lutetium (Lu), neodymium (Nd), praseodymium (Pr), promethium (Pm), samarium (Sm), terbium (Tb), thulium (Tm), ytterbium (Yb). The other two elements that are sometimes included in the wider REE category are scandium (Sc) and yttrium (Y), since they share many properties with the lanthanides and frequently occur along with them in raw deposits. These seventeen elements are essential from a technological point of view because they are employed in numerous applications. They are considered as 'critical' because there are few mining sites around the world. In fact, the supply of most REEs to the international market is dominated by China, as it hosts most of the few available mines.

Recently, a subclass of NEAs has been established, which is named Easily Recoverable Objects (ERO) \citep{2013CeMDA.116..367G}. These bodies crossing the near-Earth space are proposed as candidates for first mining tests. In general, they are defined by both being accessible by having a low relative velocity $\Delta v$, which allows approaching them using a spacecraft, and by their proximity to our planet in order to be suitable for extraction in near-Earth space-based facilities. Obviously both conditions do not consider the asteroid type and neglect the specific bulk chemical and mechanical properties that have consequences on the economic cost of the overall process.

The Moon also plays a role in the strategic expansion of space mining. It offers a nearby environment where ISRU techniques and mining technologies can be tested under reduced gravity and with fewer logistical constraints compared to asteroid missions. Extracting resources from the lunar surface to support off-Earth infrastructure could reduce dependence on expensive Earth launches. In any case, additional studies on the mechanical differences of the rock-forming minerals of lunar soils, compared with terrestrial ones are encouraged \citep{GREBOLTOMAS2025102110}.

 The future exploration of Earth’s satellite will provide opportunities to establish and test mining initiatives that can be worthwhile to be applied later in asteroids. For example, the extraction of raw materials in the Moon to build future manned bases could be cheaper than their direct transport from Earth. Obviously the extraction procedures should be developed and tested under low gravity with a minimum amount of resources and tools. In any case, establishing the lunar surface to perform the first tests is a more realistic approach than doing such experiments on a distant small asteroid exhibiting an almost zero-gravity environment. 

Significant progress in the exploration of the Moon and the asteroids is being made in parallel. During the last decades we have seen that some space agencies have successfully achieved the first sample-return initiatives from minor bodies. The two first missions: Stardust (NASA) and Hayabusa 2 (JAXA) successfully returned materials from comet 81P/Wild 2 and asteroid Itokawa, respectively. Although these pioneer missions only brought tiny amounts of material, they successfully allowed researchers to complete the characterization of their bulk compositions \citep{2006Sci...314.1711B, 2011Sci...333.1113N}.

Two additional space missions, Hayabusa 2 and OSIRIS-REx, have achieved sample return from two primitive carbonaceous asteroids, respectively 162163 Ryugu and 101955 Bennu \citep{2017SSRv..212..925L, 2024M&PS...59.2453L, 2025GeocJ..59...45Y}. Both missions were planned to characterize the nature of these asteroids, and improve our assignation of asteroidal materials to carbonaceous chondrite groups. So far the successful recovery of Hayabusa 2 of Ryugu samples has allowed establishing a genetic heritage between Ryugu and the CI group of the carbonaceous chondrites (hereafter CCs) \citep{2023NatAs...7..633P}. According to these results, asteroid Ryugu probably is not so highly affected by aqueous alteration as the CI chondrites, but it cannot be ruled out that the samples are biased to be representative of the non-hydrated outer layer \citep{2023NatAs...7..170N}. Certainly, such km-sized carbonaceous asteroid probably exemplifies the heterogeneities to be found in accretionary bodies, exhibiting similarities with other CC groups. There is clearly a need to demonstrate that the samples are representative of the materials forming the body as has been made for Hayabusa 2 \citep{2022Sci...375.1011T, 2023Sci...379.7850Y}. In a similar way, recent OSIRIS-REx results about the composition of asteroid Bennu are the key to establish its link with the CI group of carbonaceous chondrites \citep{2023Sci...379.7850Y, 2024M&PS...59.3044J, 2025NatAs.tmp..179B}.

Then, our current study of the bulk elemental compositions of several meteorite specimens of well described chemical groups can exemplify how they can be a valuable source of information for future sample-return missions. Different lines of evidence suggest that the arrival of CC materials to Earth is biased towards high-strength rocks, also affected by impacts and in some cases by aqueous alteration \citep{2006ApJ...652.1768B, 2009PASA...26..289T, 2019SSRv..215...18T}. 

Certainly, we confront a paradigm: despite we have thousands of CC specimens in our collections, these are not fully representative of all the bodies available out there. Some asteroids are not sampled by nature because they lie far from the main resonances delivering them, or their rocks, to the near-Earth region. In addition, some C-rich asteroids are probably too fragile and, when excavated by impacts, perhaps releasing only small rocks and pebbles that are in practice too small to survive atmospheric interaction \citep{2009PASA...26..289T}. Space weathering and impact processing modify the surfaces of asteroids over billions of years, often obscuring their true bulk composition \citep{2012Icar..220..466C, 2014MNRAS.437..227T}.  Understanding these alterations is essential for accurately assessing the extractable content of target bodies. Thus, it would be highly desirable to obtain better answers concerning the amount of commercially profitable material that might be found in pristine undifferentiated asteroids \citep{1997AcAau..41..637S, 2014P&SS...91...20E}.

This paper evaluates the bulk elemental compositions of various CC groups to identify the elements with the greatest potential for resource extraction. We show that undifferentiated asteroids—despite often being overlooked in favor of metal-rich differentiated bodies—offer significant economic potential for specific metals, REEs, and volatiles. We identify several chondrite groups with significant potential for mining. Our main goal is to exemplify how, in view of the bulk elemental composition analyses and recent sample-return studies, undifferentiated asteroids could also be considered as important targets for future exploration and mining, at least for selected elements. In other words, not only metal-rich differentiated asteroids may be economically profitable, as envisioned for example in \cite{1994JGR....9921129K}. In any case, our results reinforce the idea that these primordial relics of the solar system deserve greater attention in the roadmap toward sustainable space-based resource utilization. 

\section{Data reduction, theoretical approach and observations}

In order to learn about the reliability of asteroid mining we have performed bulk chemistry measurements of several meteorite specimens belonging to CC groups, most of them from the NASA Antarctic collection. To control for the effects of terrestrial alteration in finds, we compare the measured mean bulk elemental composition of Antarctic finds with a recovered fall: Orgueil. 

\begin{table*}
\centering
\caption{The carbonaceous chondrites analyzed in this work sorted by alphabetical order. We also give its chondrite group, petrologic type, the total known mass (TKW), and its recovery as fall (F) or find (f). Complete names of the listed abbreviations of Antarctic meteorites are Allan Hills (ALH), Dominion Range (DOM), Grosvenor Mountains (GRO), Larkman Nunatak (LAR), Lewis Cliff (LEW), Miller Range (MIL) and Szabo Bluff (SZA).}
\label{tab: CCs}
\begin{tabular}{lcccc}
\toprule
\textbf{Meteorite} & \textbf{Group} & \textbf{Petrologic types} & \textbf{TKW (g)} & \textbf{Fall/find (year)} \\ \midrule
Orgueil     & CI & 1.0 & 14,000 & F (1864) \\
GRO17004    & CM & 2.0 & 52.9   & f (2017)  \\
EET96029    & CM & 2.0 & 843    & f (1996)  \\
MIL13005    & CM & 1.5 & 192.7  & f (2013)  \\
LEW85311    & CM & 2.0 & 199.5  & f (1985)  \\
ALH84029    & CM & 2.0 & 119.8  & f (1984)  \\
LEW90500    & CM & 2.0 & 295    & f (1990)  \\
LAR12247    & CR & 2.0 & 137.4  & f (2012)  \\
GRO17063    & CR & 2.0 & 388    & f (2017)  \\
GRO17064    & CR & 2.0 & 226    & f (2017)  \\
DOM18319    & CO & 3.0 & 33.9   & f (2018)  \\
MIL090010   & CO & 3.0 & 2,490  & f (2009)  \\
ALH83108    & CO & 3.5 & 1519   & f (1983)  \\
DOM10104    & CO & 3.0 & 119.8  & f (1984) \\
MIL11118    & CO & 3.0 & 39.5   & f (2011)  \\
SZA12431    & CO & 3.0 & 443    & f (2012)  \\
MET01017    & CV & 3.7 & 238    & f (2001)  \\
ALH84028    & CV & 3.7 & 736    & f (1984)  \\
MIL07002    & CV & 3.7 & 758    & f (2007)  \\
LAR12002    & CV & 3.0 & 4,860  & F (2012)  \\
LAP02206    & CV & 3.7 & 1,285  & f (2002)  \\
GRA06101    & CV & 3.7 & 3,560  & f (2006)  \\
Allende     & CV & 3.0 & $2\cdot 10^6$ & F (1969) \\
GRO17059    & CK & 5.0 & 102.4  & f (2017)  \\
LAR04318    & CK & 4.0 & 53.3   & f (2003)  \\
EET16006    & CK & 5.0 & 50.2   & f (2016)  \\
GRO17169    & CK & 4.0 & 79     & f (2017)  \\
MIL090031   & Ureilite   & - & 101    & f (2009)  \\ \bottomrule
\end{tabular}
\end{table*}

The meteorites studied in this exploratory work are listed in Table \ref{tab: CCs}. Our sample set comprises petrological types ranging from 1 to 5 and includes 6 different meteorite families: CI, CO, CM, CK, CV and CR. Samples were analyzed by an ICP-AES (Inductively coupled plasma atomic emission spectroscopy) ICAP 6500 ThermoElectron for major elements (Na, Mg, Al, Si, P, K, Ca, Ti, Fe), and an ICP-MS Thermoscientific iCAP TQ for minor and trace elements.

Solutions were prepared from $\sim$0.25~g of each sample fluxed with $\sim$0.5~g of Li-metaborate and dissolved in 100~mL of 1~M HNO$_3$, together with 1~drop of HF. Five standard US Geol. Survey \footnote{\url{https://www.usgs.gov/}} reference materials were used for external calibration; while internal calibration was made using Rh as internal standard.

\begin{figure*}
    \centering
    \includegraphics[width=\textwidth]{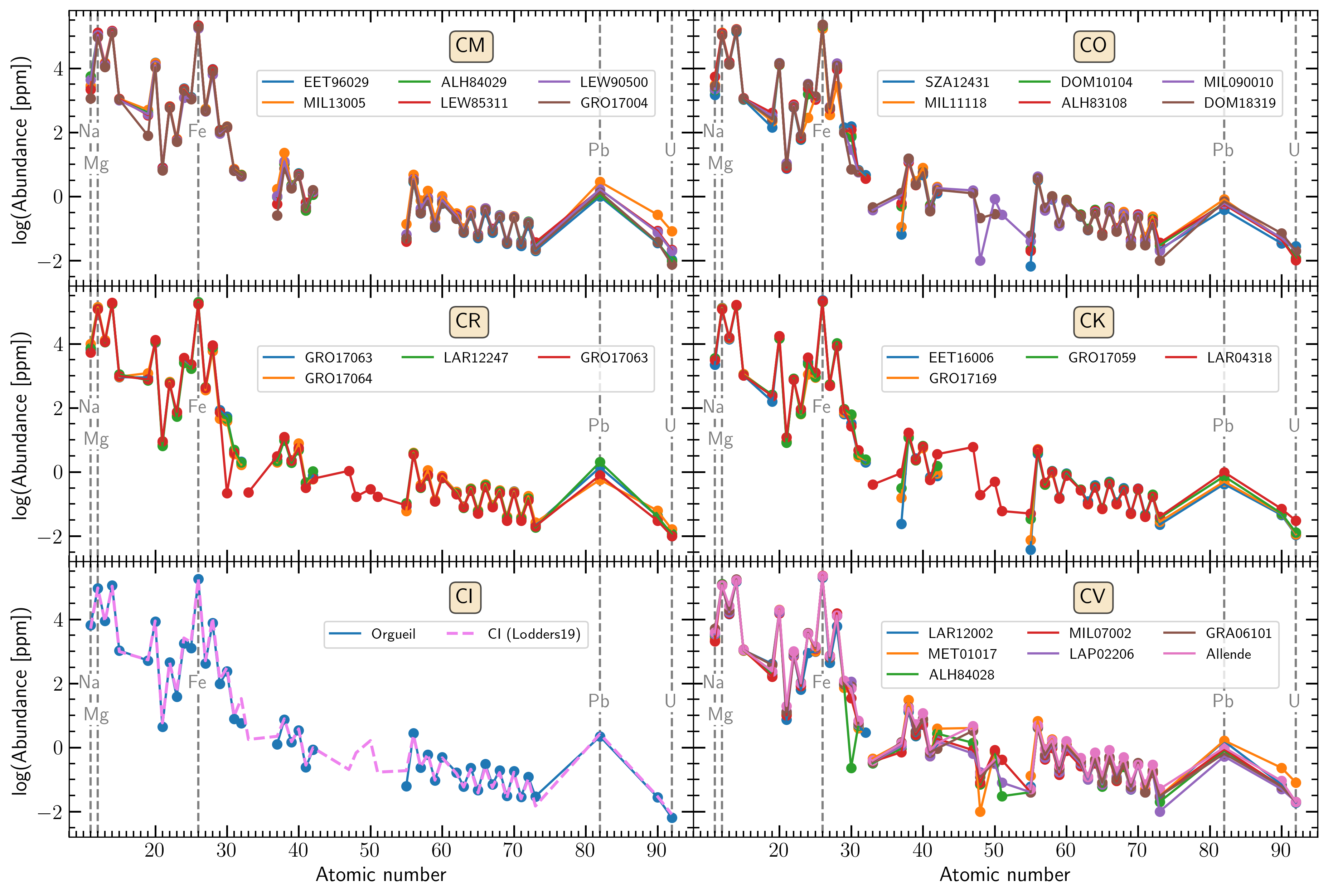}
    \caption{Bulk elemental composition of analyzed CC specimens. Vertical dashed lines mark the atomic number of relevant atomic species. In the CI panel we compared our Orgueil measurements with those of CI chondrites \citep{2019nuco.conf..165L}.}
    \label{fig: log abundances}
\end{figure*}

We successfully analyzed the elemental abundance of 46 elements between Na ($Z = 11$) and U ($Z = 92$). Due to the nature of the ICP-MS technique and sample digestion, most transition elements abundance could not be determined. As it has no known stable isotopes, the abundance of Pm ($Z = 61$) could not be determined either. We present the obtained abundances for the major and minor elements in Tables \ref{tab: abundances main} and \ref{tab: abundances min}, respectively. Additionally, we have gathered all this data in Figure \ref{fig: log abundances}. The elemental abundances of the Orgueil meteorite have been used as an indicator of the analysis quality. As seen in the corresponding panel of Figure \ref{fig: log abundances}, its elemental abundances are in accordance to those of CI reported in \citet{2019nuco.conf..165L}. Thus, our ICP-MS analysis does not suffer from any systematic error. 

Furthermore, 89 per cent of meteorites analyzed in this work are finds from the NASA Antarctic collection. Among the different sources of REE contamination, we may discard those anthropogenic due to the remoteness of the finding place \citep{2009laveuf_pedogenesisree}. Atmospheric deposition could be the most important REE compositional alteration. However, wet and dry depositions are considered minority REE sources in usual continental soils \citep[e.g.][]{2004Wang_BiogeochemicalREE}. Hence, we consider that our meteorite samples were collected without any impurity on REE due to terrestrial weathering.

\section{Results and discusion} \label{sec: results}

Next, we will discuss whether some of the CC meteorite groups may be promising ores for asteroid mining. In Section \ref{subsec: transition metals} we first focus on the transition metal abundances. In Section \ref{subsec: ree} we discuss the abundances of REE in the studied meteorites.

\subsection{Transition metals} \label{subsec: transition metals}

\begin{figure*}
    \centering
    \includegraphics[width=\textwidth]{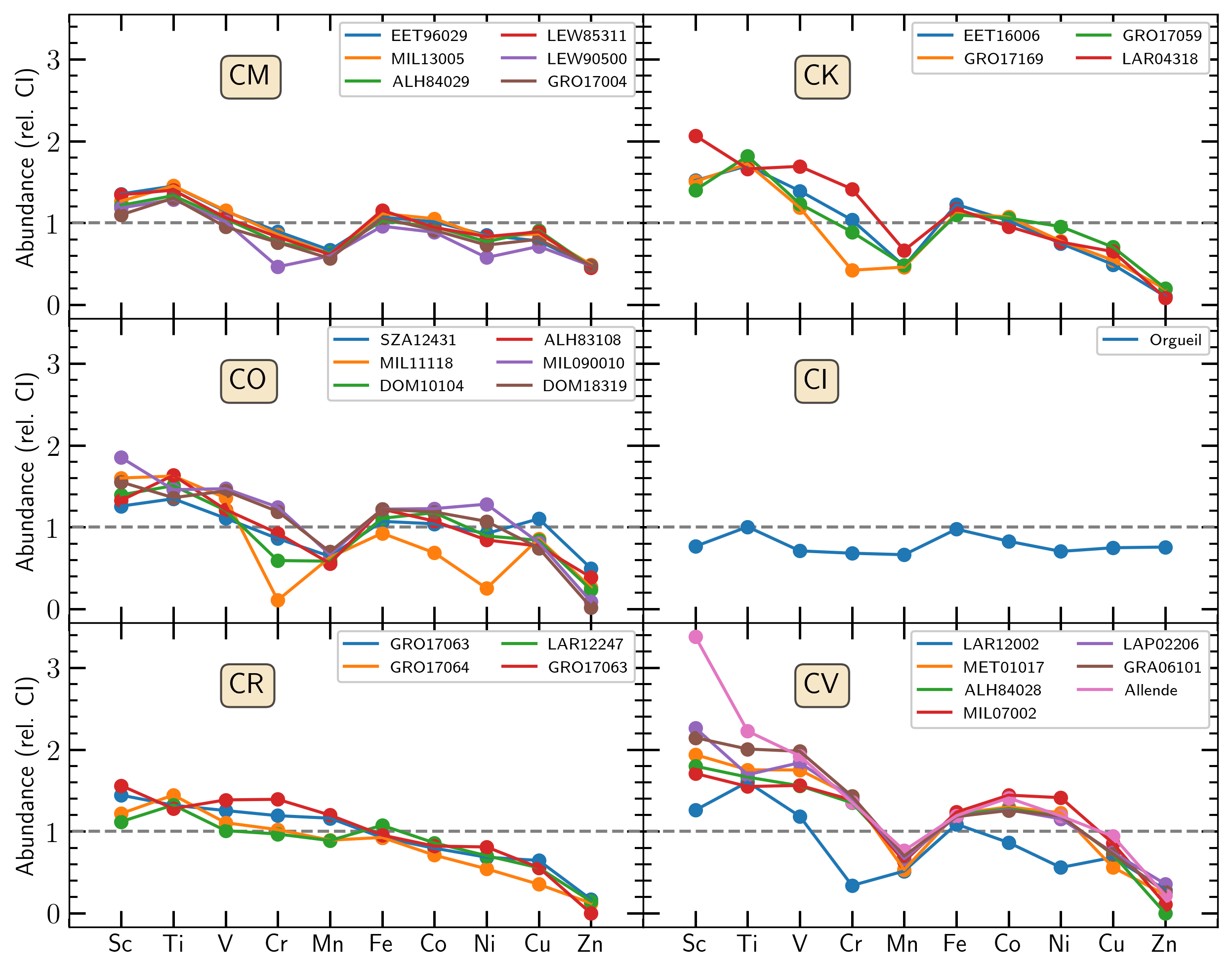}
    \caption{Measured chemical abundances of valuable metals in the selected CC specimens.}
    \label{fig: metal abundances}
\end{figure*}

We present in Figure \ref{fig: metal abundances} the elemental abundances of the third-period transition metals for the CC meteorites investigated in this work. In the figure, each meteorite has been grouped based on its specific group family. All abundances have been normalized with respect to the CI values reported in \citet{2019nuco.conf..165L}. There are clearly uniform trends within a given group of meteorites. Even though some meteorites exhibit elemental differences compared to the group trend (e.g. GRO17063 for CK or MIL11118 for CO), in our opinion these are not outliers because the general data scattering for each element is low. 

\begin{figure}
    \centering
    \includegraphics[width=\columnwidth]{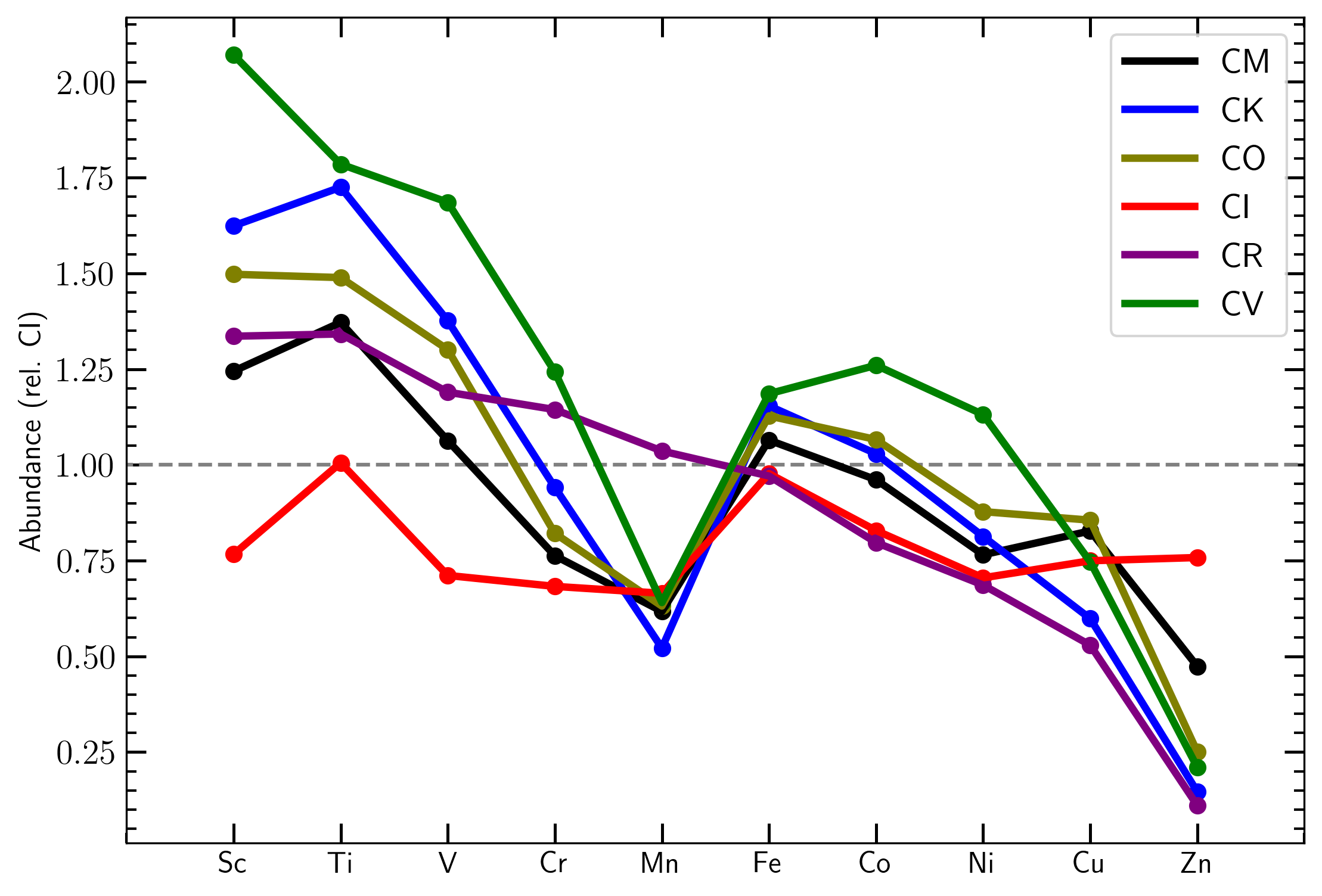}
    \caption{Mean chemical abundances of third-row transition elements in the selected CC specimens. The mean values for each group take their source from the data in Figure \ref{fig: metal abundances}. The elemental abundances are shown relative to the CI values reported in \citet{2019nuco.conf..165L}.}
    \label{fig: metal mean}
\end{figure}

In order to properly compare the different meteorite families, we have taken the mean elemental values of each group. These mean trends are shown in Figure \ref{fig: metal mean}. We have omitted the standard deviation errors in this figure to ease its interpretation. It should be noted that for CI we have only analyzed one single meteorite: Orgueil. It is remarkable the similar pattern identified for the CV and CK groups, previously presented as related to each other: CK being a collisional processed group initially associated with the CV chondrites 
\citep{2021MNRAS.507..651T}. By looking at Figure \ref{fig: metal abundances} we can notice a significant enrichment in Sc and Ti for Allende, a meteorite known for being significantly affected by shock metamorphism \citep{1994Metic..29R.509N}. The
high content found for Sc in Allende could be related with the presence of coarse-grained Ca and Al-rich Inclusions (CAIs) that 
are usually enriched in refractory siderophile and lithophile elements, particularly with high abundance in Sc \citep{2014ChEG...74..507P}.

A clear trend present in all groups is that CR, CV, CM, CK and CO are all Ti-wealthy groups compared with CI. Their abundance of Ti ranges from 1.5 to 2 times the abundance presence in CI. Based on our data, CV and CK chondrite groups appear to be the most Ti-rich samples, almost 2 times the abundance present in CI. For high atomic numbers (Co, Ni, Cu and Zn) there is a decrease on the elemental abundance, compared to those on CI, which is present in all meteorite families. It is worth noting that, among all the meteorite families, CR appear to be specially Mn-rich (about 1.5 times the abundance found on CI chondrites), while the other groups are deficient on Mn, compared with CIs. It is also noticeable in Figure \ref{fig: metal mean} that Cu and particularly Zn are depleted compared with CIs. 

\begin{figure*}
    \centering
    \includegraphics[width=\textwidth]{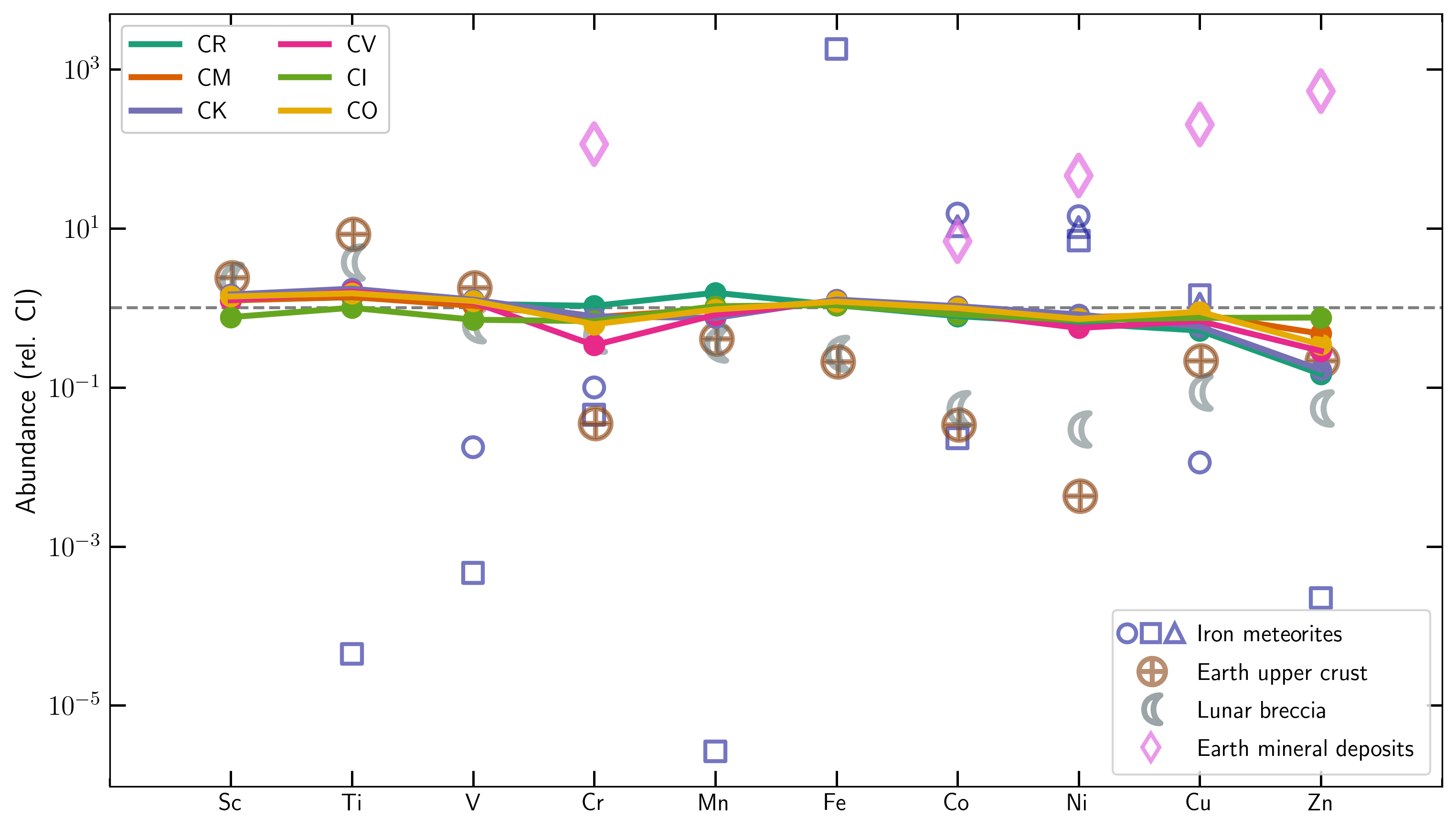}
    \caption{Mean chemical abundances of the third-period transition elements in the analyzed charbonaceous chondrites compared to other bodies in the Solar System. All abundances have been normalized to the reported CI values in \citet{2019nuco.conf..165L}. Bluish markers (circles, squares and triangles) show the relative abundance of transition elements in iron meteorites: \citet{2005GeCoA..69.4733C, 2014Chernonozhkin_iron, 2014DuanRegelous_iron}, respectively. In \citet{2005GeCoA..69.4733C} several iron meteorites abundances are reported. In this figure we arbitrarily show that of the Cape of Good Hope IVB iron meteorite to ease the interpretation, as all iron meteorites from the work show similar abundances. The upper Earth crust elemental abundances from \citet{2003TrGeo...3....1R} are shown as browny circles with meridian lines. Pink diamonds show the relative elemental abundance in Earths most proliferant mineral deposits: Bushveld complex (South Africa) for Cr \citep{2015Cawthorn_Bushveld}, Tenke Fungurume mines (D. R. Congo) for Co and Cu \citep{2012Fay_TenkeFungurume, 2014LundinMiningCorp_TenkeFungurume}, Ranglan deposit (Canada) for Ni \citep{2004Seabrook_Raglan} and the Red Dog mine (Alaska, USA) for Zn \citep{2002Jennings_RedDog, 2004Kelley_RedDog}. Crescent grey moons show the relative abundance values from the PCA 02007 lunar meteorite \citep[lunar breccia,][]{2010M&PS...45..917J}. In \citet{2010M&PS...45..917J} the analysis of other breccias are shown, but they all show similar elemental compositions. As in the case of iron meteorites, we only show the abundance of one of them to ease the graph interpretation.}
    \label{fig: metal comparison}
\end{figure*}

We think that the asteroids can be a good supply for certain transition elements, which may be useful in future space missions. To justify our claim, in Figure \ref{fig: metal comparison} we compare the mean elemental abundances from our chondrite groups (Figure \ref{fig: metal mean}) with the reported abundances in several Solar System surfaces. In the scale of this figure, the relative abundances in the analyzed meteorites are quite uniform. This is due to the fact that all of the analyzed meteorites are carbonaceous chondrites. Hence, when set in a comparison context, they tend to group together and appear as similar bodies. Compared with iron meteorites \citep{2005GeCoA..69.4733C, 2014Chernonozhkin_iron, 2014DuanRegelous_iron}, CCs have higher abundances of most transition elements. 

In particular, the CH chondrite group contains significant amounts of Fe and Ni \citep{2016M&PS...51.1795M}. Unfortunately, few specimens of that chondrite group are known and most have experienced extensive terrestrial weathering, so they are not included in our study. 

One of the main features of iron meteorites is the large presence of Fe-Ni-alloys (kamacite and taenite, \citet{1974mcp..book.....W}). Thus, it is not surprising that they stand out as Fe-rich bodies: about 1000 times the Fe present in CIs. Additionally, they also feature higher amounts of Co and Ni than CCs. Regarding the rest of transition metal elements, CCs are from 10 to $10^5$ times richer than iron meteorites. This might make CCs an important source of transition metal elements such as Ti, V, Cr, Mn and Zn. 

Compared to the abundances on Earth upper crust \citep{2003TrGeo...3....1R}, CCs appear to have lower abundances only for low-$Z$ transition metal elements (Sc, Ti and V). Since Earth suffered from a differentiation process, heavier elements (such as Fe and Ni) are more common in the innermost layers of the planet. On the other hand, CCs have not undergone these processes due to their small size  \citep{2006mess.book..679B}. Without this differentiation process, high-$Z$ elements are more common in CCs than in the upper crust of Earth. However, this higher abundance is significantly lower to that found in Earth's main mineral deposits \citep[e.g.][]{2015Cawthorn_Bushveld, 2012Fay_TenkeFungurume, 2004Seabrook_Raglan, 2004Kelley_RedDog}, which is a consequence of different geological process that concentrate the critical elements. Based on this comparison, mining a CC parent asteroid can be more expensive than extracting minerals from the Earth. In terms of relative abundance, some iron meteorites show similar Co and Ni compositions than the main mineral deposits. 

Finally, in Figure \ref{fig: metal comparison} we show the third-row transition metal abundances for lunar breccias. Lunar breccia compositions are quite similar to those in Earth upper crust, as expected from its geological origin and its differentiation processes \citep{2014pacs.book..213W}. Then, the compositional trends are similar to those discussed above: while the lunar surface is low-$Z$ rich, it lacks of higher $Z$ transition metal elements. Besides, the Moon has no known intrinsic mineral deposits from where to extract these elements. Based on the results of Figure \ref{fig: metal comparison}, impact craters where fragments of the projectiles became implanted forming breccias could be significant mining resources for future missions. Instead of bringing metals from Earth, these minerals could be extracted in-situ from the lunar surface. In any case, the study of the parent asteroids of chondrites might identify if there is any kind of evolutionary process that allows for the enrichment of any of these elements. Our results suggest that, given the porous nature of CCs, collisional gardening and brecciation could favour the enrichment in some of these elements by implantation. In any case, given the chemical heterogeneity found for some specimens of a same group, this should be studied case by case due to the stochastic nature of the magnitude and composition of the projectiles.

\subsection{Rare-Earth elements} \label{subsec: ree}

The relevance of the relatively scarce rare-earths is daily exemplified in the context of their direct use in technological applications. Before going any further, it is important to remark that some chemistry textbooks define the rare-earth elements comprised by the lanthanoids ($Z = 57 - 71$), scandium ($Z=21$) and yttrium ($Z=39$). In order to simplify the following discussion, we do not consider Sc and Y as REE due to the notorious chemical differences with lanthanoids, which are the major number of REE.

\begin{figure*}
    \centering
    \includegraphics[width=\textwidth]{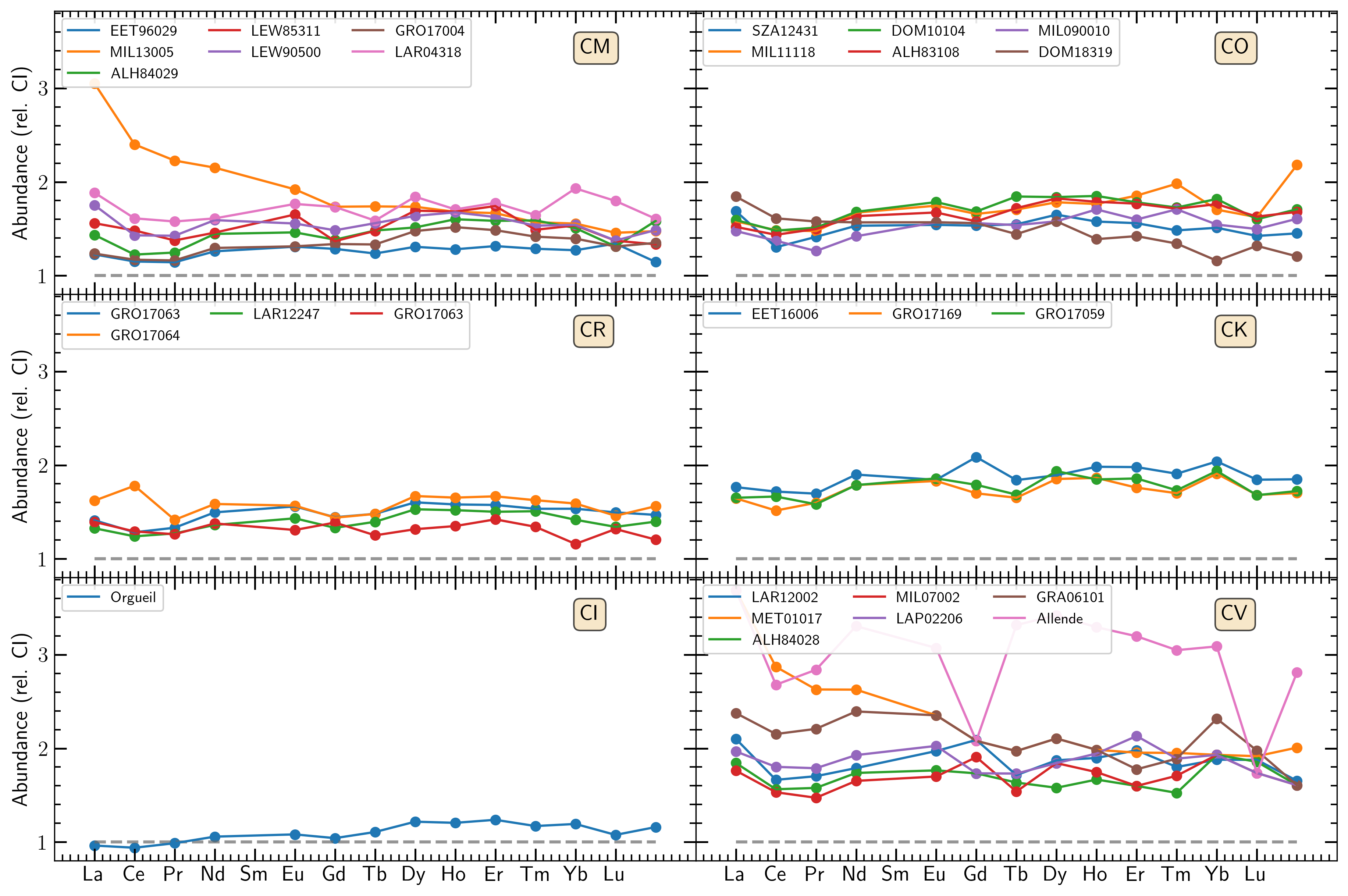}
    \caption{Measured chemical abundances of REEs in the analyzed CC specimens. Elemental abundances are normalized by the corresponding abundances in CI \citep{2019nuco.conf..165L}.}
    \label{fig: ree abundances}
\end{figure*}

We show in Figure \ref{fig: ree abundances} the measured REE abundances for each analyzed meteorite, normalized by the corresponding abundances found in CI \citep{2019nuco.conf..165L}, in a similar fashion as with transition metal elements in Figure \ref{fig: metal abundances}. Again, we see that meteorites within a same family behave similarly. Generally, there is a low scattering between abundances in a given meteorite group, excepting some particular outliers. For example, in the CR family the GRO17064 meteorite show a high cerium abundance compared with other group members, while sharing similar abundances of the other REE. A particular meteorite is MIL13005, a CM, which shows particularly high abundances for low-$Z$ REE. However, we consider this to be a peculiarity of the meteorite and not representative of the overall CM group, since the other meteorites have similar abundances among them.

We observe in general trends that some CC groups are between 1.5 to 2 times the abundances measured for the CI group. In particular, the CV, CK and CO groups contain enough key elements to be considered valuable materials to mine. 

\begin{figure}
    \centering
    \includegraphics[width = \columnwidth]{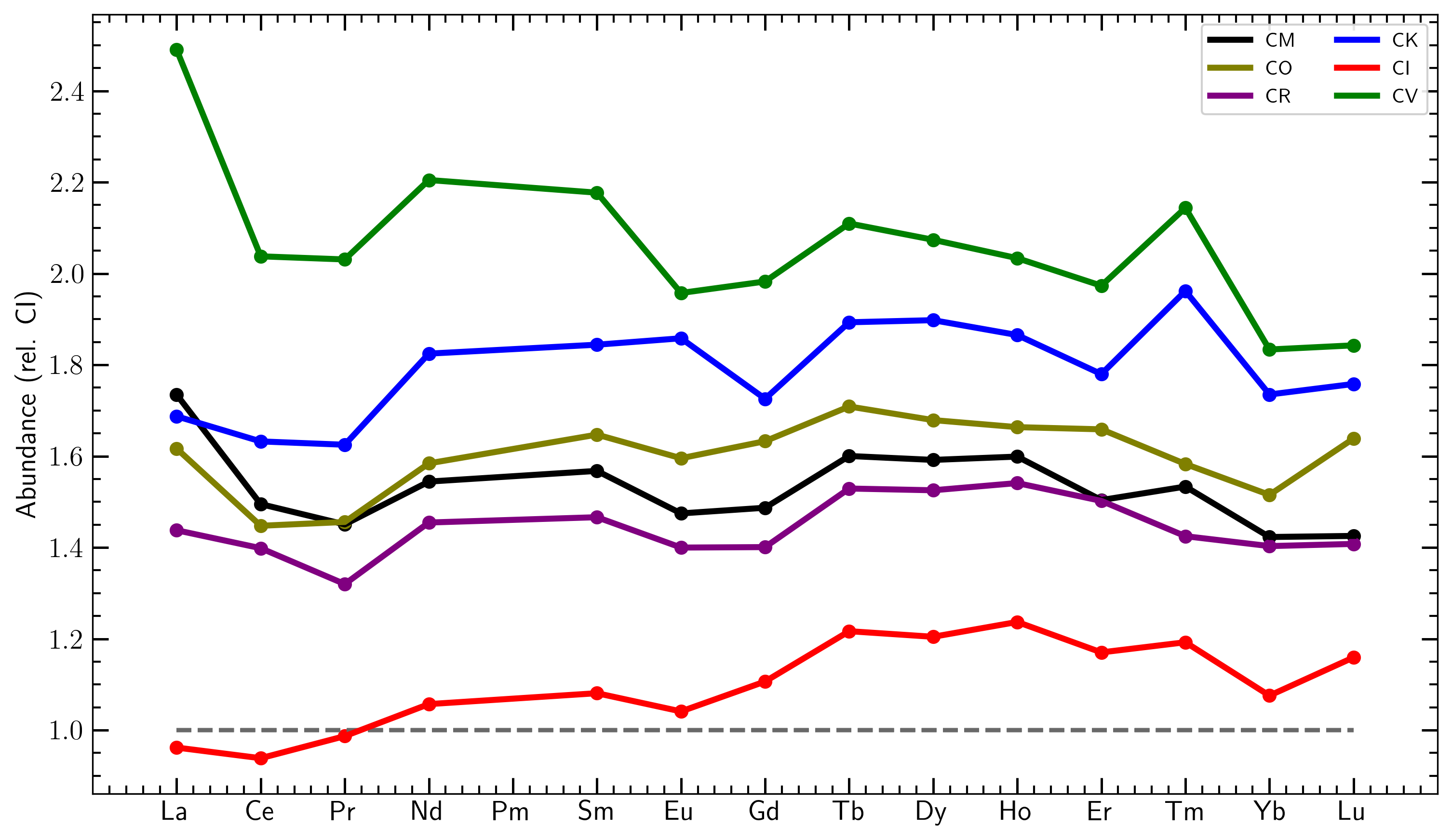}
    \caption{Mean REE abundance for each meteorite group. Abundances are normalized with respect to the reported CI abundances in \citet{2019nuco.conf..165L}.  Standard deviations are not shown in order to ease the interpretation of the figure. As hinted by Figure \ref{fig: ree abundances}, they are low and do not affect the drawn conclusions.}
    \label{fig: ree mean abundances}
\end{figure}

The REE elemental abundances on the analyzed samples are properly compared in Figure \ref{fig: ree mean abundances}. This figure includes the mean abundances in each group, similar to transition elements in the section 3.1. The plot has been simplified by not showing the standard deviation errors. As seen in Figure \ref{fig: ree abundances} the trends within a given group are quite uniform, meaning that the errors are not as high to prevent comparing the REE abundances only with the mean values. 

A particularity of Figure \ref{fig: ree mean abundances} is that it appears to be a clear stratification in the plot based on the meteorite group. The abundance patterns in terms of atomic number hardly ever intersect. The CK and CV families are the richest in all REE, followed by CO, CM and CR. The CI family host the lowest amount of REE among CCs. This stratification did not occur when comparing the transition metal elements abundances (Figure \ref{fig: metal mean}), which could be related with the chemistry of lanthanides.

This stratification also suggests that the total amount of REE (particularly lanthanoids) can be related with the petrologic type of a particular chondrite group. We further explore this assumption in Section \ref{subsubsec: ree petrologic}.

Lanthanoids are mostly present in nature in its trivalent form (Ln$^{+3}$, where Ln denotes any lanthanoid). However, cerium ($Z = 58$) and europium ($Z = 63$) can appear in other forms. In reducing environments, Eu$^{3+}$ can be reduced to Eu$^{2+}$. This divalent ion is similar in both size and charge to the Ca$^{2+}$ ion, which makes is suitable to fit on the cation vacancies on plagioclases. Hence, due to lattice substitution, plagioclases can be doped with europium. We suspect that positive anomalies could be related with the accumulation of Eu in primary plagioclase, particularly subjected to collisional processing, producing heat and metasomatism in the rock forming minerals. For example, Eu anomalies found in terrestrial granitic rocks were interpreted as evidence of earlier separation of a mineral phase such as plagioclase as consequence of thermal processing during differentiation as demonstrated \citet{1983GeCoA..47.1131F}.

The previous work conducted by \citet{2021MNRAS.507..651T} analyzed the reflectance spectra of CV and CK carbonaceous chondrites. They concluded that both meteorite families indeed show similar properties. In fact, they suggest that both families origin could be in the same parent body, with CK being the highly-collided CV counterpart. The fact that these two families are related is endorsed by the REE distribution in Figure \ref{fig: ree mean abundances}. Besides of having the highest REE relative abundances, both meteorite families show a positive Eu anomaly, whereas all other families have a negative Eu anomaly. This similarity supports that CK and CV may have a common origin. 

\begin{figure*}
    \centering
    \includegraphics[width = \textwidth]{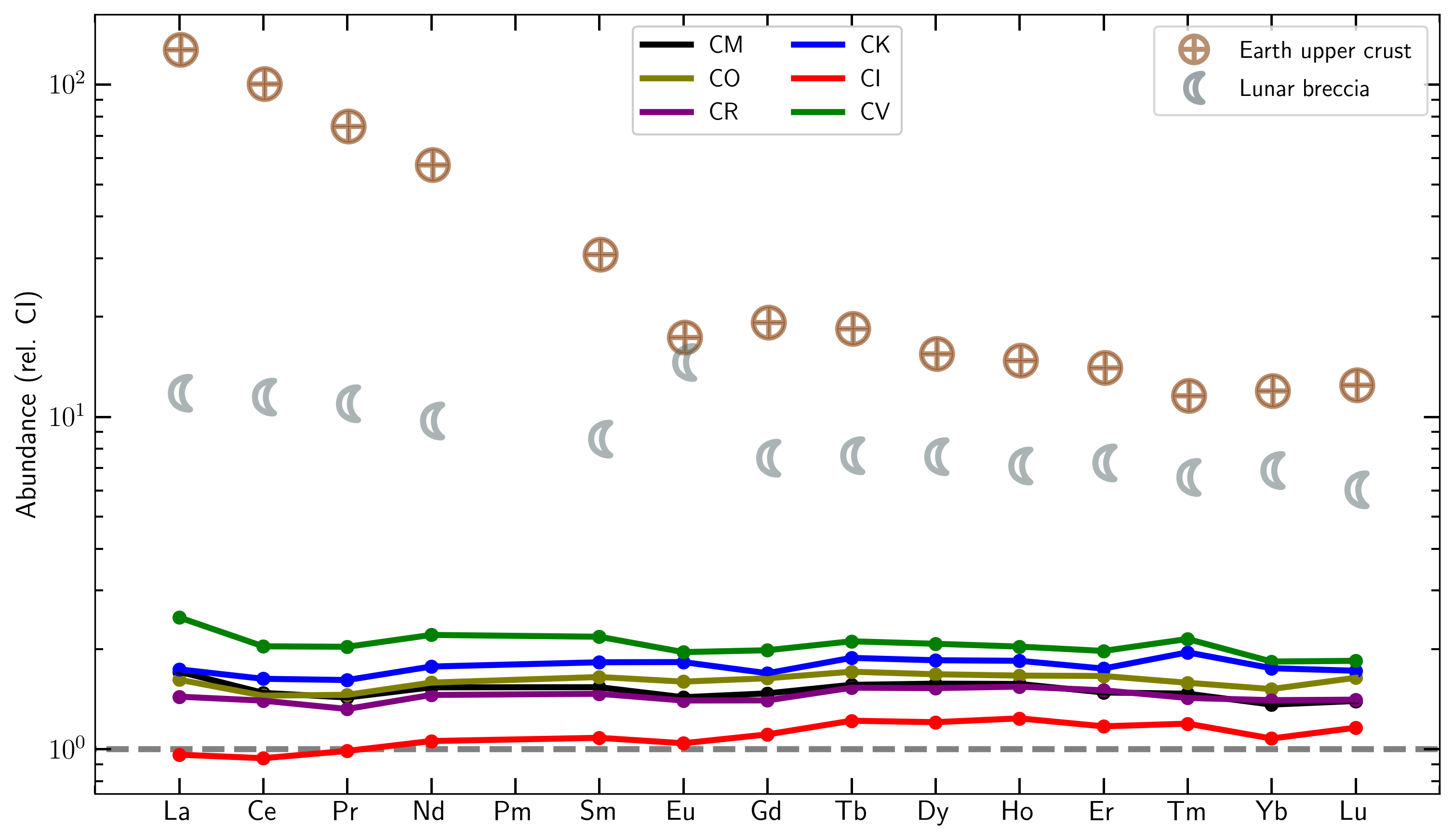}
    \caption{Comparison of the mean REE relative abundance in the analyzed meteorite families with other bodies in the Solar System. All abundances have been normalized to the reported CI values in \citet{2019nuco.conf..165L}. As in Figure \ref{fig: metal comparison}, Earth upper crust abundances are extracted from \citet{2003TrGeo...3....1R}. Lunar breccia abundances result from the LA-ICP-MS analysis of the PCA 02007 lunar meteorite \citep{2010M&PS...45..917J}.}
    \label{fig: ree comparison}
\end{figure*}

In Figure \ref{fig: ree comparison} we compare the REE abundance in our meteorites with those found in the Earth upper crust \citep{2005GeCoA..69.3419R} and in lunar breccia meteorites \citep{2010M&PS...45..917J}. In this case, both Earth and Moon seem to host larger REE abundances compared to carbonaceous chondrites. On the bulk Earth upper crust there is from 10 to 100 times more REEs than on CCs. Based on the results in Figure \ref{fig: ree comparison}, investing in mining to extract REE from CCs parent asteroids seems not worthy. Even in future lunar population missions CCs would not be a great choice to extract REE, much depending in the development of new techniques to solve the challenges behind their extraction.

An interesting feature shown in Figure \ref{fig: ree comparison} is that the Earth upper crust has a slight negative Eu anomaly, indicative of a fractionized magma solidification, where plagioclases would be found in lower layers. On the other hand, PCA 02007 lunar meteorite is feldspathic, probably formed from regoliths from lunar feldspathic regions. It has a notorious positive Eu anomaly, which is also present in other lunar feldspathic meteorites \citep{2010M&PS...45..917J}. On the other hand, the basaltic regolith from \citet{2010M&PS...45..917J} features a negative Eu anomaly much like Earth upper crust. It means that Eu is depleted relative to the other REEs in most carbonaceous chondrite groups, with exception of the CK (see Figure 6).

\subsubsection{Dependency of the total REE abundance with the petrologic type} \label{subsubsec: ree petrologic}

The apparent stratification shown in Figure \ref{fig: ree mean abundances} intriguingly suggest a relation between the REE elemental abundances and the meteorite group. In turn, meteorite families are deeply related with the petrologic type, which is a proxy for the degree of aqueous alteration or thermal metamorphism that the meteorite has suffered \citep{2005CRPhy...6..303D}.

\begin{figure}
    \centering
    \includegraphics[width=\columnwidth]{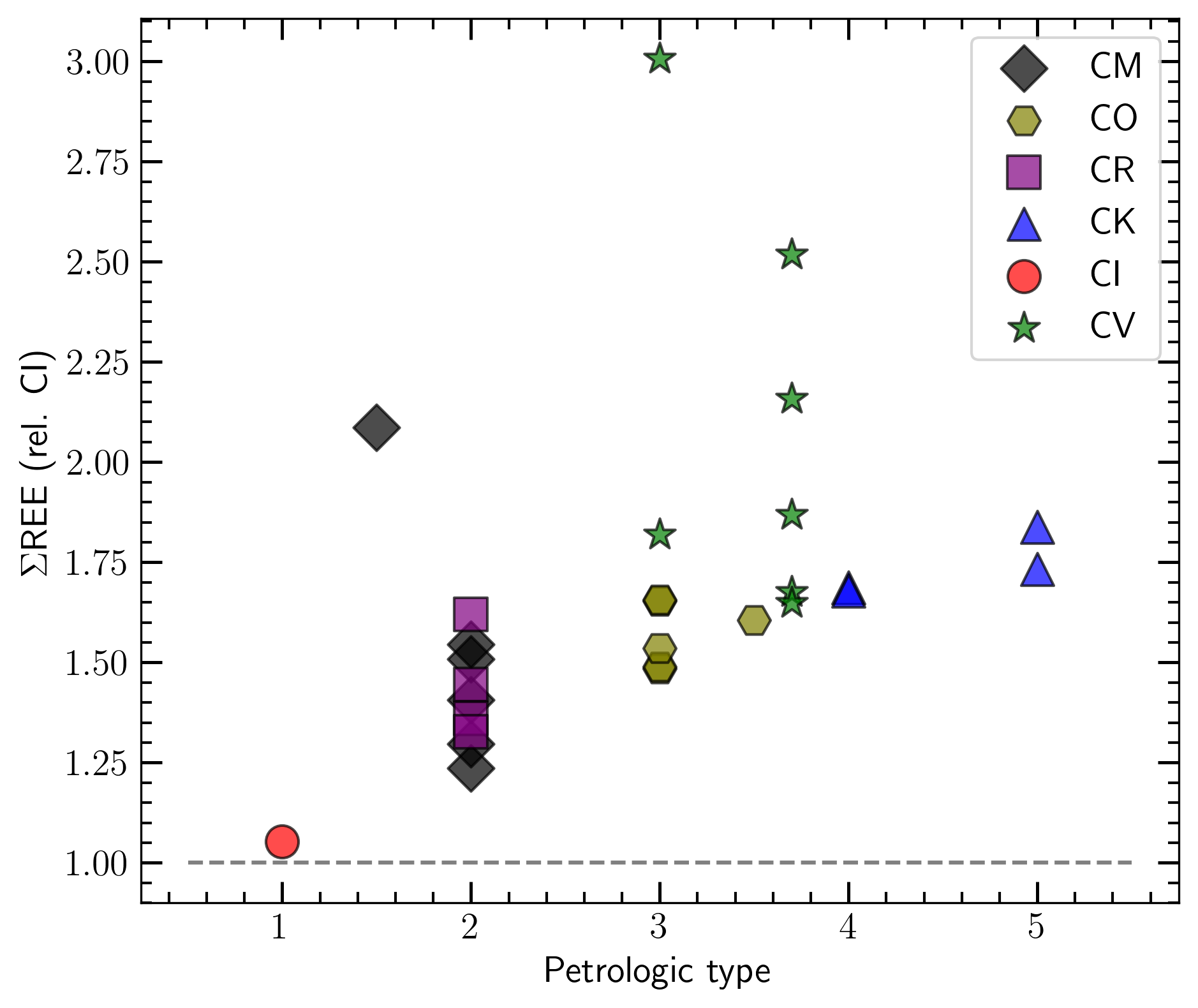}
    \caption{Relation between the total REE abundance in a meteorite and the petrologic type. Rare-earth elements only include lanthanides. Each point in the plot accounts for a single meteorite. The $y$ values are the ratio between the total abundance of REE in the meteorite and the total abundance of REE in CIs given by \citet{2019nuco.conf..165L}. Petrologic type values are those in Table \ref{tab: CCs}.}
    \label{fig: ree petrologic}
\end{figure}

In order to assess this issue we compare in Figure \ref{fig: ree petrologic} the total relative REE abundance with the petrologic type of the meteorites. The values on the $y$ axis are found as dividing the total abundance of REE on a meteorite by the total abundance of REE on CIs \citep{2019nuco.conf..165L}. As shown in Figure \ref{fig: ree mean abundances}, aqueous alteration appears to decrease the content in REEs, perhaps reflecting preaccretionary processes \citep{1998M&PS...33.1113B}. A petrologic type of 3 is usually considered a pristine chondrite, relatively unaffected by thermal processing or aqueous alteration. In general, our results for these specimens indicate abundances about 1.6 times those found for the CI group. CV and CM chondrite groups seem to host higher abundance of REE, being so heterogeneous that this could be consequence of the implantation of foreign materials during collisional gardening (e.g. many CM chondrites are breccias \citep{2006GeCoA..70.1271T}). On the other hand, it is difficult to extract a conclusion for the higher petrologic degree measured in the CK group as the values are slightly over the ones found for petrologic type 3.

Our results indicate that CCs contain valuable materials and that the non-aqueously altered groups are particularly hosting them in native form, suggesting them as primary targets for exploration and mining. In particular, the most pristine CO chondrites, typically exhibiting less thermal processing, have a much higher abundance of REE than ordinary chondrites \citep{1989AMR.....2..279E} or the CI chondrite group usually taken as a chemical reference \citep{2010ASSP...16..379L}.

However, the CR and CM chondrites experienced parent body aqueous alteration to different degrees, often associated with collisional compaction
\citep{2021M&PS...56..546V}. Despite of the collisional processing, they can still contain significant abundance of key elements for mining, particularly because many are breccias that could retain certain materials from the foreign projectiles
\citep{2007GeCoA..71.2361R}. In any case, the meteorite evidence points that the water was heterogeneously accreted on the parent asteroids of both groups, and its availability at cm-scale could be highly variable \citep{2019SSRv..215...18T}. In those cases in which water was locally abundant, it promoted the mobilization of soluble elements during aqueous alteration, and the ulterior formation of secondary minerals, but perhaps not bulk chemical changes. Obviously, the previous has the exception of transitional asteroids being extremely affected by impacts and water flow, and ending their days as dormant comets (rich in CI-like altered materials). Compared with the extensively aqueously altered, and partially depleted CI group, the parent bodies of CR and CM might be also worthwhile to mine.

\section{Main implications to identify ore-bearing asteroids}

In view of our findings, with the objective of identifying bodies that are worth mining, we should be putting our efforts in the study of pristine asteroids. Then, we need to match our meteorites with their parent asteroids, a really non-trivial task. The best way to characterize remotely asteroids is using the Bus and Bus-DeMeo taxonomies \citep{2009Icar..202..160D}. Asteroids doesn't reflect the light arrived from the Sun as a mirror due to the different ability of rock-forming minerals in their surfaces to reflect the sunlight. The concept of spectral class is inferred remotely from shape, slope and observed features in asteroid reflectance spectra. Asteroids classified in a given spectral class are related with each other, due to share common reflectance spectra, and having similar composition and mineralogy \citep{2009Icar..202..160D}. 

Promising ore-bearing candidates are asteroids associated with the K spectral class, often associated with the CO and CV groups \citep{1988Metic..23..256B, 2009Icar..202..119C, 2024PSJ.....5..194B}. Their reflectance spectra exhibit marked olivine and spinel absorption bands like those found in CO chondrites \citep{1997AcAau..41..637S, 2014MNRAS.437..227T}. To distinguish the provisional assignment between the CO and CV chondrites and the K-class of asteroids, we strongly believe that a sample return mission from the 221 Eos family could have great scientific impact and much economic potential. 

In turn, the Cg spectral class reflectance features seem to be pointing towards a common chondritic reservoir for the CV–CK chondrite groups as demonstrated in previous works \citep{2013GeCoA.108...45W, 2021MNRAS.507..651T}. These two chondrite groups are similar in chemical grounds, but exhibit a different degree of collisional evolution in their parent asteroids. Probably being the CK highly shocked fragments of a larger CV progenitor could explain the affinities found. 

Something to keep in mind is the required advance in developing new techniques, and efficient steps to implement them under space conditions. This is clearly exemplified by the complex extraction of REEs, taking into account the usual processes at work on Earth. REEs are part of natural minerals, often forming complex mixtures. Then, the REEs extraction involves chemical separation, a challenging process, not only for the processes required but also for the environmental issues concerning the products and residues involved in chemical separations. 

In Table A.3 we see two results for GRO 17063.34 and GRO 17063.7 corresponding to two different chips analyzed of the same meteorite. For a same meteorite, in this case GRO 17063, the digits provided by NASA after the dot allow us to identify the exact location of the chip analyzed before its extraction from the meteorite specimen. We include both values as a test to visualize the data scattering.

We also made some intrinsic tests with paired specimens to find out our analytical accuracy. Paired meteorites are specimens located close to each other that are associated with a same meteorite fall (i.e., they are the consequence of the disruption in the atmosphere of the progenitor meteoroid). An example is the similar elemental values found for GRO 17063 and GRO 17064, both CR2 chondrites probably associated with a same fall. Other meteorites that might be paired a priori, but which are found to belong to different groups, turn out to be clearly distinguishable. This is the case, for instance, of the CK5 chondrite GRO 17059 and the CR2 chondrite GRO 17063.

According to the elemental abundances found in this paper, perhaps we should not only consider metal-rich differentiated bodies as source of metal resources \citep{2014P&SS...91...20E}. Conversely, we should also recognize that other chondrite groups can be  interesting for mining, but probably we will need additional sample-return missions to find out their parent asteroids \citep{2017ASSP...46...73M}. The OSIRIS-REx and Hayabusa 2 missions provided new clues about the composition of carbonaceous asteroids, their inherited volatile content \citep{2023A&A...671A...2D, 2025NatAs...9..199G}, and even their connection with another bodies \citep{2025Natur.637.1072M}. In addition, future sample return missions are expected to achieve ground truth about the nature of potentially hazardous asteroids (PHAs) and extinct comets in near-Earth space \citep{2020A&A...641A..58T}.

In general we should expect that the chondritic meteorites arrived to Earth are good proxies of the rock-forming materials forming asteroids. It is particularly true when a fast recovery of meteorites arrived to Earth's surface is achieved because it allows minimizing the degradative effects of terrestrial alteration \citep{2025MNRAS.542.1743G}. In fact, we have studied the mechanical properties of Itokawa asteroid finding that their constituent minerals have similar mechanical behaviour than those forming ordinary chondrites \citep{2019A&A...629A.119T}. In any case, the exploitation of ordinary chondrites will depend on the degree of annealing, brecciation and shock compaction and lithification experienced by as consequence of collisional processing \citep{2004GeCoA..68..673R}. Such processes are minimized for some groups of carbonaceous chondrites, due to their highly porous primordial nature \citep{2006ApJ...652.1768B}.

By increasing the number of sample-return missions we will be able to establish accurate cosmochemical links between undifferentiated asteroids and the chondrite groups in our meteorite collections. The potential of such missions will be relevant for: 1) The identification of the parent asteroid of CC groups providing key scientific information concerning the accretion of planetesimals with many implications in Cosmochemistry, 2) The identification of a relative fragile target to test mining procedures and subsequent sample return at a massive scale, perhaps with dedicated spacecrafts at circumlunar orbit, and 3) assessment of the physical properties of the bodies in the framework of planetary defense \citep{2022air..book.....T}

However, inferred abundances of many rare (minority) metals deserve further discussion (Figure \ref{fig: metal abundances}). The abundance of rare metals could be explained in an evolutionary context. First, one should bear in mind that CO chondrites are considered very pristine meteorites that experienced little aqueous alteration or metamorphism. ALHA77307 is, for example, one of the most pristine meteorites known. We think that the reason for such differences in the abundance of valuable elements, compared to other ordinary or CC groups, is the relative absence of thermal metamorphism and aqueous alteration experienced by the parent body of CO chondrites (COs) that enables them to preserve their native metals. Water is relatively absent in the CO group members, so we should certainly not look for the liquid element in the CO parent asteroid. 

OSIRIS-REx results identified asteroid Bennu in an intermediate position along the aqueous alteration continuum, among the extensively altered type 1 and the less-altered petrologic type 2 or 3 carbonaceous chondrites \citep{2025NatAs.tmp..179B}. In general, the lessons learned from sample return missions demonstrate that every single undifferentiated asteroid might be highly heterogeneous, and having experienced different degrees of aqueous alteration. In particular, our analytical results are consistent with significant variability in La, Y, Rb, and Ce, among other elements, probably as consequence of parent body alteration processes. In that context, for aqueously altered CCs, we could expect that they experienced catalytic processes as a function of the degree of alteration with significant consequences for the bulk chemistry and the production of complex organics \citep{2016SciRep..6..38888, 2021A&A...650A.160C}.

Given that our paper is focused in quantifying the abundance of metals and REEs, we have excluded from our discussion the intrinsic relevance of water, abundant in some CC groups \citep{2017RSPTA.37550384A, 2018SSRv..214...36A, 2019SSRv..215...18T}, but we remark that if the liquid element is the target resource, the aqueously altered parent bodies of CI and CM groups should be explored. The low thermal inertia of their surfaces make these asteroids able to retain a significant percent of water in mass, although it might be difficult to extract due to be mainly bounded within hydrated minerals \citep{2018E&PSL.482...23M, 2019SSRv..215...18T, 2023M&PS...58.1760L}. To infer from remote observations the best asteroid candidates, the presence of water-soluble sulfate and phosphate-bearing minerals in Bennu, give us significant clues \citep{2025NatAs.tmp..179B}. In addition, the valuable Hayabusa 2 data will allow us a comparative study of the elemental abundances of Ryugu compared with the CI group, to figure out its idoneity as a proxy of the solar composition \citep{2025GeocJ..59...45Y}.

Consequently, in view of our analytical results on valuable elements, we think those future projects to achieve asteroid mining in rocky asteroids should be focused in the study of pristine asteroids, probably of the K spectral class. This latter assignment is inferred from the study of reflectance spectra and by noticing that K-class asteroids exhibit marked olivine and spinel absorption bands like those found in CO and CV chondrites \citep{2009Icar..202..119C, 2012Icar..220..466C, 2014MNRAS.437..227T}. New sample-return missions from asteroids are encouraged as they will probably provide new clues on other relevant candidates.  

\section{Conclusions}

Human civilization is becoming increasingly aware that the exploitation of terrestrial resources is approaching its limits. The concept of indefinite growth is unsustainable and has contributed to the current global ecological and economic crises. In response, space mining is emerging as a promising avenue, with private companies and national agencies exploring its feasibility. The Moon will likely serve as a testing ground for new extraction and processing technologies, paving the way for more ambitious operations on asteroids.

Eventually, selecting asteroid targets that offer both scientific and economic value could represent a mutual beneficial approach. However, to make this a reality, new international policies must be implemented to regulate the exploration and utilization of minor planetary bodies in a sustainable and equitable manner. We are entering a remarkable era of solar system exploration, characterized by multiple active missions and successful sample returns. These missions are beginning to provide long-awaited “ground truth” data on asteroid compositions. Nevertheless, millions of unvisited bodies remain, accessible only through the study of meteorites delivered to Earth by collisions and dynamic processes over millions of years. Care must be taken when extrapolating meteorite data to entire parent bodies, as delivery biases and surface weathering may distort our understanding of their original composition.

Although metallic asteroids are often prioritized for future mining operations, undifferentiated, carbonaceous asteroids may also offer significant value due to their content of volatiles, organics, and critical elements. In this study, we analyzed the bulk elemental composition of various carbonaceous chondrite groups and found that relatively pristine types, particularly the CO chondrites, may be of interest for resource extraction. Other CC groups, such as CM, CI, and CR, may also hold potential under specific technological and mission conditions, assuming that meteorite and sample-return materials are representative of their parent bodies. Remote sensing technologies, particularly reflectance and thermal infrared spectroscopy, will continue to play a key role in the preliminary assessment and prioritization of asteroid mining targets.

Based on our results, we propose the following key conclusions:

a) Pristine, undifferentiated asteroids that likely correspond to the K spectral class could be prioritized for detailed study and potential mining. This classification is based on reflectance spectra showing characteristic olivine and spinel absorption bands, similar to those found in CO chondrites.

b) If water is the target resource, aqueously altered asteroids should be explored. These are likely the parent bodies of CI, CM, and CR chondrites, which often display OH$^-$ absorption features in near-infrared spectra. However, meteorite evidence indicates that aqueous alteration may be spatially heterogeneous within these bodies, requiring in situ studies.

c) The parent bodies of CO and CR chondrites should be identified as a matter of priority. These groups appear not only scientifically important but also economically viable for future resource extraction missions.

d) As Earth's population continues to grow and transitions toward clean energy technologies, the demand for critical elements such as copper (Cu), cobalt (Co), nickel (Ni), and manganese (Mn) is expected to increase. This growing demand has already fueled geopolitical tensions. Space-based resource acquisition may help alleviate these pressures in the future.

e) We found that the total amount of REE (particularly lanthanoids) can be associated with the petrologic type of a chondrite group. CV and CM chondrite groups seem to host higher abundance of REE.

f) A comprehensive geochemical study of carbonaceous chondrites is essential for identifying promising targets for space mining. However, this effort must be coupled with new sample-return missions to verify the parent body identities. Given the observed compositional heterogeneity within CC groups, detailed in-situ exploratory missions will also be required to locate and characterize pristine regions suitable for extraction.

g) The development of space mining will depend heavily on advancements in autonomous robotics, in-situ resource utilization, and remote geochemical sensing. These technologies must be adapted to the unique conditions of low-gravity and structurally complex asteroid environments. In-situ resource utilization is also a key enabler for future long-duration missions to the Moon and Mars, reducing dependence on Earth-based resupply.

h) As we move forward, it is critical to incorporate environmental and ethical guidelines into space resource utilization in order to avoid repeating the unsustainable practices of Earth-based extraction. Clear regulatory frameworks and transparent operational standards should be prioritized from the outset.

i) Finally, the scale and complexity of asteroid mining require international collaboration. Cooperative frameworks will help ensure equitable access, peaceful use, and shared scientific and economic returns from space resource activities.

In addition to economic and scientific motivations, a better understanding of asteroid composition is also crucial for planetary defense strategies, particularly for modeling impact scenarios and designing effective deflection techniques. The pursuit of asteroid resources intersects with multiple long-term goals in planetary science, sustainability, human exploration, and planetary defense. The successful integration of geochemical knowledge, technological innovation, and international governance will define our capacity to engage in the exploration of the Solar System responsibly and effectively.

\section*{Acknowledgments}
	JMT-R and JII thank the MEC for AYA research grant PID2021-128062NB-I00 (FEDER/Ministerio de Ciencia e Innovación y Agencia Estatal de Investigación) as well as the Spanish program Unidad de Excelencia María de Maeztu CEX2020-001058-M. PG-T acknowledges the financial support from the Spanish MCIU through the FPI predoctoral fellowship PRE2022-104624. This work is part of the doctoral thesis of PG-T (Doctoral Program in Physics at Universitat Autònoma de Barcelona). MG acknowledges the Academy of Finland project no. 325806 (PlanetS). The programme of development within Priority-2030 is acknowledged for supporting the research at UrFU. Finally, NASA Meteorite Working Group and Johnson Space Center meteorite curators are acknowledged for providing the Antarctic carbonaceous chondrites. We also want to express our sincere gratitude for the valuable effort made over the years in the recovery of Antartic samples by ANSMET (The Antarctic Search for Meteorites program).  

\section*{Data Availability}

All the analytical data generated in the course of this study is compiled in the Appendix A as Table A.1. An Excel compilation of the data is available under request.


\bibliographystyle{mnras}
\bibliography{main} 



\appendix

\section{ICP-MS measured abundances} \label{app: measures abundances tables}
\begin{table*}
\caption{ICP-MS elemental abundance results for the main elements in each meteorite specimen analyzed. Values given in weight percentage (wt\%).}
\label{tab: abundances main}
\begin{tabular}{lcccccccccc}
\toprule
\textbf{Meteorite} & \textbf{Group} & \textbf{Na} & \textbf{Mg} & \textbf{Al} & \textbf{Si} & \textbf{P} & \textbf{K} & \textbf{Ca} & \textbf{Ti} & \textbf{Fe} \\ \midrule
Orgueil      & CI & 0.89 & 15.32 & 1.68 & 24.12 & 0.24 & 0.06 & 1.18 & 0.08 & 25.93 \\
GRO17004.7   & CM & 0.15 & 15.35 & 2.10 & 27.67 & 0.24 & 0.01 & 1.49 & 0.10 & 28.13 \\
EET96029.97  & CM & 0.36 & 17.51 & 2.29 & 29.67 & 0.24 & 0.04 & 2.07 & 0.11 & 28.41 \\
MIL13005.23  & CM & 0.39 & 17.60 & 2.84 & 30.61 & 0.26 & 0.06 & 2.01 & 0.11 & 29.59 \\
LEW85311.91  & CM & 0.29 & 21.27 & 2.59 & 31.31 & 0.26 & 0.04 & 1.74 & 0.11 & 30.65 \\
ALH84029.71  & CM & 0.77 & 18.16 & 2.31 & 28.31 & 0.24 & 0.05 & 1.60 & 0.10 & 27.24 \\
LEW90500.91  & CM & 0.59 & 18.15 & 2.42 & 28.85 & 0.23 & 0.04 & 1.69 & 0.10 & 25.50 \\
LAR12247.18  & CR & 0.99 & 20.54 & 2.10 & 37.84 & 0.25 & 0.09 & 1.67 & 0.10 & 28.53 \\
GRO17063.34  & CR & 0.72 & 19.90 & 2.20 & 41.28 & 0.23 & 0.09 & 1.82 & 0.10 & 25.36 \\
GRO17063.7   & CR & 0.89 & 20.93 & 2.25 & 38.90 & 0.21 & 0.11 & 1.53 & 0.10 & 24.57 \\
GRO17064.7   & CR & 1.34 & 22.94 & 2.46 & 40.08 & 0.22 & 0.15 & 1.60 & 0.11 & 24.60 \\
DOM18319.6   & CO & 0.37 & 18.01 & 2.47 & 33.30 & 0.26 & 0.03 & 2.00 & 0.10 & 32.32 \\
MIL090010.61 & CO & 0.30 & 18.39 & 2.55 & 33.85 & 0.26 & 0.04 & 2.04 & 0.11 & 32.39 \\
ALH83108.123 & CO & 0.74 & 20.42 & 2.94 & 34.68 & 0.27 & 0.05 & 1.87 & 0.12 & 32.39 \\
DOM10104.30  & CO & 0.28 & 19.93 & 2.70 & 32.57 & 0.25 & 0.03 & 1.80 & 0.11 & 29.43 \\
MIL11118.5   & CO & 0.41 & 21.52 & 2.95 & 35.53 & 0.26 & 0.03 & 1.93 & 0.12 & 24.60 \\
SZA12431.9   & CO & 0.20 & 18.34 & 2.52 & 29.01 & 0.24 & 0.02 & 2.05 & 0.10 & 28.44 \\
MET01017.51  & CV & 0.33 & 20.59 & 3.31 & 35.11 & 0.24 & 0.02 & 2.79 & 0.13 & 32.30 \\
ALH84028.171 & CV & 0.41 & 21.10 & 2.90 & 35.54 & 0.25 & 0.02 & 2.38 & 0.12 & 31.93 \\
MIL07002.58  & CV & 0.28 & 19.78 & 2.80 & 35.09 & 0.26 & 0.02 & 2.17 & 0.12 & 32.81 \\
LAR12002.48  & CV & 0.48 & 19.94 & 2.73 & 32.75 & 0.25 & 0.05 & 2.39 & 0.12 & 28.92 \\
LAP02206.93  & CV & 0.46 & 19.71 & 3.11 & 36.13 & 0.26 & 0.03 & 2.33 & 0.13 & 31.31 \\
GRA06101.111 & CV & 0.68 & 19.69 & 3.68 & 37.26 & 0.25 & 0.05 & 2.71 & 0.15 & 31.37 \\
Allende      & CV & 0.52 & 19.03 & 5.00 & 35.03 & 0.25 & 0.03 & 2.68 & 0.17 & 31.63 \\
GRO17059.6   & CK & 0.47 & 20.52 & 2.96 & 32.81 & 0.24 & 0.03 & 2.03 & 0.14 & 29.14 \\
LAR04318.42  & CK & 0.42 & 19.98 & 2.98 & 35.37 & 0.23 & 0.03 & 2.44 & 0.12 & 31.12 \\
EET16006.6   & CK & 0.30 & 22.00 & 2.59 & 33.71 & 0.23 & 0.02 & 2.08 & 0.13 & 32.56 \\
GRO17169.6   & CK & 0.43 & 21.47 & 2.83 & 34.62 & 0.26 & 0.03 & 2.38 & 0.13 & 29.67 \\
MIL090031.15 &  Ureilite  & 0.03 & 27.45 & 0.18 & 38.33 & 0.08 & b.d.l. & 0.96 & 0.02 & 19.43 \\ \bottomrule
\end{tabular}
\end{table*}

\begin{table*}
\caption{ICP-MS elemental abundances for the minor elements in each meteorite specimen analyzed. Values given in ppm. We also indicate which elements are below detection limits (b.d.l.)}
\label{tab: abundances min}
\resizebox{\textwidth}{!}{%
\begin{tabular}{lc|cccccccccccccccccc}
\toprule
\textbf{Meteorite} & \textbf{Group} & \textbf{Sc} & \textbf{V} & \textbf{Cr} & \textbf{Co} & \textbf{Ni} & \textbf{Cu} & \textbf{Zn} & \textbf{Ga} & \textbf{Rb} & \textbf{Sr} & \textbf{Y} & \textbf{Zr} & \textbf{Nb} & \textbf{Mo} & \textbf{Cs} & \textbf{Ba} & \textbf{La} & \\ \midrule
Orgueil & CI & 4.47 & 38.12 & 1781.66 & 420.88 & 7719.25 & 97.44 & 235.72 & 7.73 & 1.26 & 7.37 & 1.46 & 3.41 & 0.24 & 0.85 & 0.06 & 2.76 & 0.23 & \\
EET96029.97 & CM & 7.90 & 61.24 & 2338.85 & 515.53 & 9319.21 & 101.46 & 147.59 & 6.56 & 1.00 & 11.17 & 2.06 & 5.31 & 0.47 & 1.49 & 0.05 & 2.74 & 0.30 & \\
MIL13005.23 & CM & 7.39 & 61.82 & 2242.99 & 534.91 & 9051.51 & 112.85 & 153.37 & 7.23 & 1.71 & 22.75 & 2.28 & 5.06 & 0.36 & 1.36 & 0.14 & 4.82 & 0.74 & \\
ALH84029.71 & CM & 7.09 & 56.36 & 1995.12 & 485.48 & 8446.67 & 118.17 & 144.41 & 6.73 & 0.96 & 10.43 & 2.19 & 5.00 & 0.37 & 1.14 & 0.04 & 3.02 & 0.35 & \\
LEW85311.91 & CM & 7.84 & 57.09 & 2163.94 & 480.22 & 9135.66 & 115.88 & 140.38 & 6.68 & 0.58 & 12.09 & 1.95 & 4.89 & 0.65 & 1.30 & 0.04 & 3.47 & 0.38 & \\
LEW90500.91 & CM & 6.90 & 54.44 & 1213.61 & 450.64 & 6328.20 & 92.86 & 148.59 & 6.26 & 1.01 & 12.21 & 2.01 & 4.57 & 0.49 & 1.39 & 0.06 & 3.44 & 0.43 & \\
GRO17004.7 & CM & 6.41 & 51.02 & 1990.22 & 462.88 & 7993.04 & 104.18 & 149.55 & 6.39 & 0.26 & 7.48 & 1.78 & 4.41 & 0.43 & 1.60 & 0.05 & 3.07 & 0.30 & \\
GRO17063.34 & CR & 9.10 & 74.28 & 3635.16 & 417.33 & 8878.62 & 72.72 & 0.22 & 3.97 & 3.11 & 12.24 & 2.25 & 5.54 & 0.32 & 0.61 & 0.09 & 3.51 & 0.34 & \\
GRO17063.7 & CR & 8.42 & 67.31 & 3113.79 & 404.16 & 7506.58 & 83.96 & 52.90 & 4.42 & 2.36 & 10.94 & 2.30 & 5.43 & 0.43 & 0.76 & 0.08 & 3.73 & 0.34 & \\
GRO17064.7 & CR & 7.11 & 59.34 & 2668.74 & 361.08 & 5955.62 & 46.12 & 38.58 & 3.56 & 2.00 & 11.53 & 2.18 & 7.86 & 0.46 & 0.60 & 0.06 & 3.96 & 0.40 & \\
LAR12247.18 & CR & 6.53 & 54.17 & 2528.96 & 437.14 & 7708.99 & 72.36 & 45.57 & 4.91 & 2.18 & 9.65 & 1.93 & 4.71 & 0.47 & 1.04 & 0.10 & 3.77 & 0.32 & \\
DOM18319.6 & CO & 9.05 & 77.38 & 3111.07 & 602.88 & 11736.36 & 97.01 & 7.02 & 6.05 & 1.29 & 15.27 & 2.30 & 5.41 & 0.35 & 1.59 & 0.06 & 3.71 & 0.45 & \\
MIL090010.61 & CO & 10.79 & 78.84 & 3251.63 & 623.59 & 14035.12 & 105.60 & 28.64 & 5.78 & 1.08 & 13.32 & 2.41 & 5.38 & 0.44 & 1.81 & 0.04 & 4.17 & 0.36 & \\
ALH83108.123 & CO & 7.74 & 65.17 & 2417.45 & 546.08 & 9240.74 & 100.17 & 120.75 & 5.69 & 0.61 & 11.87 & 2.29 & 5.62 & 0.47 & 1.46 & 0.02 & 4.00 & 0.37 & \\
DOM10104.30 & CO & 8.14 & 64.69 & 1547.10 & 598.23 & 9778.64 & 109.28 & 73.65 & 5.73 & 0.49 & 13.82 & 2.39 & 5.65 & 0.54 & 1.55 & 0.02 & 3.83 & 0.39 & \\
MIL11118.5 & CO & 9.33 & 72.82 & 283.34 & 349.39 & 2778.97 & 111.86 & 82.27 & 6.10 & 0.11 & 11.53 & 3.03 & 7.86 & 0.52 & 1.97 & 0.02 & 3.52 & 0.39 & \\
SZA12431.9 & CO & 7.33 & 59.51 & 2247.94 & 529.27 & 10104.44 & 143.40 & 154.65 & 6.66 & 0.06 & 15.19 & 2.38 & 4.63 & 0.37 & 1.26 & 0.01 & 3.14 & 0.41 & \\
LAR12002.48 & CV & 7.35 & 63.56 & 886.97 & 439.11 & 6132.51 & 88.93 & 89.43 & 4.44 & 1.30 & 12.86 & 2.23 & 5.21 & 0.67 & 1.22 & 0.06 & 4.10 & 0.51 & \\
MET01017.51 & CV & 11.30 & 93.88 & 3731.08 & 659.87 & 13452.75 & 72.96 & 71.08 & 3.95 & 1.36 & 30.1 & 2.80 & 7.42 & 0.55 & 3.85 & 0.13 & 6.59 & 0.90 & \\
ALH84028.171 & CV & 10.48 & 83.53 & 3523.92 & 650.18 & 12988.89 & 93.36 & 0.23 & 4.80 & 0.93 & 15.64 & 2.45 & 5.64 & 0.55 & 2.69 & 0.04 & 4.56 & 0.45 & \\
MIL07002.58 & CV & 9.96 & 83.74 & 3612.95 & 733.58 & 15468.56 & 111.22 & 34.53 & 5.22 & 0.71 & 15.22 & 2.50 & 5.29 & 0.54 & 1.85 & 0.04 & 4.39 & 0.43 & \\
LAP02206.93 & CV & 13.20 & 98.63 & 3679.95 & 643.43 & 12640.76 & 96.60 & 111.68 & 5.19 & 1.1 & 15.0 & 3.12 & 7.55 & 0.54 & 1.56 & 0.04 & 4.31 & 0.48 & \\
GRA06101.111 & CV & 12.51 & 106.10 & 3735.52 & 638.38 & 13005.75 & 94.28 & 81.31 & 5.65 & 1.46 & 18.07 & 3.16 & 7.74 & 0.70 & 0.92 & 0.04 & 4.14 & 0.58 & \\
Allende & CV & 19.70 & 103.03 & 3539.81 & 716.09 & 13025.79 & 122.28 & 68.92 & 6.86 & 1.34 & 17.18 & 5.23 & 11.72 & 0.83 & 1.35 & 0.05 & 4.85 & 0.90 & \\
GRO17059.6 & CK & 8.18 & 66.13 & 2323.18 & 537.62 & 10468.42 & 91.66 & 63.18 & 3.52 & 0.31 & 11.74 & 2.31 & 6.49 & 0.58 & 1.55 & 0.03 & 4.27 & 0.40 & \\
GRO17169.6 & CK & 8.79 & 63.70 & 1104.53 & 544.92 & 8527.29 & 70.84 & 59.72 & 2.91 & 0.15 & 12.83 & 2.33 & 5.60 & 0.59 & 0.91 & 0.01 & 5.17 & 0.40 & \\
LAR04318.42 & CK & 12.05 & 90.69 & 3693.13 & 486.48 & 8397.32 & 84.54 & 26.41 & 4.76 & 0.92 & 16.82 & 2.46 & 6.16 & 0.57 & 3.57 & 0.05 & 4.92 & 0.46 & \\
EET16006.6 & CK & 8.86 & 74.59 & 2705.45 & 522.06 & 8214.24 & 64.11 & 32.41 & 3.14 & 0.02 & 12.38 & 2.61 & 5.90 & 0.70 & 0.75 & b.d.l. & 3.70 & 0.43 & \\
MIL090031.15 & Ureilite & 6.47 & 73.13 & 3752.32 & 95.13 & 1184.51 & 11.01 & 157.76 & 1.56 &  &  & 0.21 & 0.54 & 0.01 & 0.33 &  &  & 0.01 & \\ \midrule
\textbf{Meteorite} & \textbf{Family} & \textbf{Ce} & \textbf{Pr} & \textbf{Nd} & \textbf{Sm} & \textbf{Eu} & \textbf{Gd} & \textbf{Tb} & \textbf{Dy} & \textbf{Ho} & \textbf{Er} & \textbf{Tm} & \textbf{Yb} & \textbf{Lu} & \textbf{Hf} & \textbf{Ta} & \textbf{Pb} & \textbf{Th} & \textbf{U} \\ \midrule
Orgueil & CI & 0.59 & 0.09 & 0.50 & 0.17 & 0.06 & 0.23 & 0.05 & 0.30 & 0.07 & 0.19 & 0.03 & 0.18 & 0.03 & 0.12 & 0.03 & 2.20 & 0.03 & 0.01 \\
EET96029.97 & CM & 0.72 & 0.11 & 0.59 & 0.20 & 0.07 & 0.26 & 0.05 & 0.32 & 0.07 & 0.21 & 0.03 & 0.22 & 0.03 & 0.12 & 0.02 & 0.98 & 0.04 & 0.01 \\
MIL13005.23 & CM & 1.50 & 0.21 & 1.02 & 0.29 & 0.10 & 0.36 & 0.07 & 0.42 & 0.09 & 0.26 & 0.04 & 0.24 & 0.04 & 0.16 & 0.02 & 2.84 & 0.27 & 0.08 \\
ALH84029.71 & CM & 0.77 & 0.12 & 0.68 & 0.22 & 0.08 & 0.31 & 0.06 & 0.40 & 0.09 & 0.26 & 0.04 & 0.22 & 0.04 & 0.16 & 0.03 & 1.10 & 0.04 & 0.01 \\
LEW85311.91 & CM & 0.93 & 0.13 & 0.69 & 0.25 & 0.08 & 0.31 & 0.06 & 0.42 & 0.10 & 0.24 & 0.04 & 0.23 & 0.03 & 0.15 & 0.04 & 1.63 & 0.08 & 0.02 \\
LEW90500.91 & CM & 0.90 & 0.14 & 0.75 & 0.24 & 0.09 & 0.32 & 0.06 & 0.42 & 0.09 & 0.25 & 0.04 & 0.23 & 0.04 & 0.16 & 0.03 & 1.66 & 0.08 & 0.02 \\
GRO17004.7 & CM & 0.73 & 0.11 & 0.61 & 0.20 & 0.08 & 0.28 & 0.06 & 0.38 & 0.08 & 0.23 & 0.04 & 0.22 & 0.03 & 0.15 & 0.02 & 1.33 & 0.04 & 0.01 \\
GRO17063.34 & CR & 0.81 & 0.12 & 0.65 & 0.20 & 0.08 & 0.26 & 0.05 & 0.34 & 0.08 & 0.22 & 0.03 & 0.22 & 0.03 & 0.13 & 0.02 & 0.78 & 0.03 & 0.01 \\
GRO17063.7 & CR & 0.80 & 0.13 & 0.71 & 0.24 & 0.08 & 0.31 & 0.06 & 0.40 & 0.09 & 0.25 & 0.04 & 0.25 & 0.04 & 0.16 & 0.02 & 1.38 & 0.04 & 0.01 \\
GRO17064.7 & CR & 1.12 & 0.13 & 0.75 & 0.24 & 0.08 & 0.31 & 0.06 & 0.42 & 0.09 & 0.27 & 0.04 & 0.24 & 0.04 & 0.18 & 0.03 & 0.57 & 0.06 & 0.02 \\
LAR12247.18 & CR & 0.78 & 0.12 & 0.64 & 0.22 & 0.08 & 0.29 & 0.06 & 0.38 & 0.08 & 0.25 & 0.04 & 0.22 & 0.03 & 0.15 & 0.02 & 2.02 & 0.04 & 0.01 \\
DOM18319.6 & CO & 1.01 & 0.15 & 0.74 & 0.24 & 0.09 & 0.30 & 0.06 & 0.35 & 0.08 & 0.22 & 0.03 & 0.22 & 0.03 & 0.14 & 0.01 & 0.69 & 0.07 & 0.02 \\
MIL090010.61 & CO & 0.86 & 0.12 & 0.67 & 0.24 & 0.09 & 0.32 & 0.06 & 0.43 & 0.09 & 0.28 & 0.04 & 0.25 & 0.04 & 0.15 & 0.02 & 0.61 & 0.05 & 0.02 \\
ALH83108.123 & CO & 0.90 & 0.14 & 0.77 & 0.26 & 0.09 & 0.36 & 0.07 & 0.45 & 0.10 & 0.28 & 0.05 & 0.27 & 0.04 & 0.18 & 0.04 & 0.56 & 0.05 & 0.01 \\
DOM10104.30 & CO & 0.93 & 0.14 & 0.79 & 0.27 & 0.10 & 0.38 & 0.07 & 0.47 & 0.10 & 0.28 & 0.05 & 0.27 & 0.04 & 0.19 & 0.03 & 0.61 & 0.05 & 0.01 \\
MIL11118.5 & CO & 0.93 & 0.14 & 0.79 & 0.27 & 0.10 & 0.35 & 0.07 & 0.44 & 0.10 & 0.32 & 0.04 & 0.27 & 0.05 & 0.24 & 0.03 & 0.81 & 0.05 & 0.01 \\
SZA12431.9 & CO & 0.82 & 0.13 & 0.72 & 0.24 & 0.09 & 0.32 & 0.06 & 0.40 & 0.09 & 0.24 & 0.04 & 0.24 & 0.04 & 0.13 & 0.02 & 0.38 & 0.03 & 0.03 \\
LAR12002.48 & CV & 1.04 & 0.16 & 0.84 & 0.30 & 0.12 & 0.36 & 0.07 & 0.48 & 0.11 & 0.30 & 0.05 & 0.31 & 0.04 & 0.17 & 0.03 & 1.55 & 0.07 & 0.02 \\
MET01017.51 & CV & 1.80 & 0.25 & 1.24 & 0.36 & 0.12 & 0.41 & 0.08 & 0.50 & 0.11 & 0.32 & 0.05 & 0.32 & 0.05 & 0.19 & 0.03 & 1.64 & 0.23 & 0.08 \\
ALH84028.171 & CV & 0.98 & 0.15 & 0.82 & 0.27 & 0.10 & 0.34 & 0.06 & 0.42 & 0.09 & 0.25 & 0.05 & 0.31 & 0.04 & 0.17 & 0.02 & 0.79 & 0.06 & 0.02 \\
MIL07002.58 & CV & 0.96 & 0.14 & 0.78 & 0.26 & 0.11 & 0.32 & 0.07 & 0.44 & 0.09 & 0.28 & 0.05 & 0.29 & 0.04 & 0.16 & 0.03 & 0.88 & 0.06 & 0.02 \\
LAP02206.93 & CV & 1.13 & 0.17 & 0.91 & 0.31 & 0.10 & 0.36 & 0.07 & 0.49 & 0.12 & 0.31 & 0.05 & 0.29 & 0.04 & 0.18 & 0.01 & 0.52 & 0.05 & 0.02 \\
GRA06101.111 & CV & 1.35 & 0.21 & 1.13 & 0.36 & 0.12 & 0.41 & 0.08 & 0.50 & 0.10 & 0.31 & 0.06 & 0.33 & 0.04 & 0.18 & 0.03 & 0.64 & 0.06 & 0.02 \\
Allende & CV & 1.68 & 0.27 & 1.56 & 0.47 & 0.12 & 0.69 & 0.13 & 0.83 & 0.18 & 0.50 & 0.08 & 0.29 & 0.07 & 0.29 & 0.05 & 0.99 & 0.09 & 0.02 \\
GRO17059.6 & CK & 1.04 & 0.15 & 0.84 & 0.28 & 0.10 & 0.35 & 0.07 & 0.47 & 0.10 & 0.28 & 0.05 & 0.28 & 0.04 & 0.20 & 0.04 & 0.73 & 0.05 & 0.01 \\
GRO17169.6 & CK & 0.95 & 0.15 & 0.84 & 0.28 & 0.10 & 0.34 & 0.07 & 0.47 & 0.10 & 0.28 & 0.05 & 0.28 & 0.04 & 0.19 & 0.03 & 0.50 & 0.05 & 0.01 \\
LAR04318.42 & CK & 1.01 & 0.15 & 0.76 & 0.27 & 0.10 & 0.33 & 0.07 & 0.43 & 0.10 & 0.27 & 0.05 & 0.30 & 0.04 & 0.17 & 0.04 & 0.98 & 0.07 & 0.03 \\
EET16006.6 & CK & 1.08 & 0.16 & 0.90 & 0.28 & 0.12 & 0.38 & 0.07 & 0.50 & 0.11 & 0.31 & 0.05 & 0.31 & 0.05 & 0.18 & 0.02 & 0.42 & 0.05 & 0.01 \\
MIL090031.15 & Ureilite  & 0.03 & b.d.l. & 0.01 & b.d.l. & b.d.l. & 0.02 & b.d.l. & 0.04 & 0.01 & 0.03 & 0.01 & 0.05 & 0.01 & 0.01 & b.d.l. & 0.29 & b.d.l. & b.d.l. \\ \bottomrule
\end{tabular}
}
\end{table*}



\bsp	
\label{lastpage}
\end{document}